\documentclass[twocolumn,showpacs,showkeys,amssymb,nobibnotes,superscriptaddress,aps,pra,bibnotes]{revtex4-1}

\usepackage{graphicx}
\usepackage{dcolumn}
\usepackage{bm}
\usepackage[T1]{fontenc}
\usepackage{mathptmx}
\usepackage{mathrsfs}
\usepackage{physics}
\usepackage{simplewick}
\usepackage{mathrsfs}
\usepackage{amsthm,amsmath,amssymb}

\usepackage[bottom]{footmisc}

\usepackage{amssymb}
\usepackage{amsmath}
\usepackage{amsfonts}
\usepackage{bm}
\usepackage{graphicx}
\usepackage{subfigure}
\usepackage[dvipsnames]{xcolor}
\usepackage{calc}
\usepackage{epsfig}
\usepackage{color}
\usepackage[colorlinks,citecolor=blue]{hyperref}
\begin{document}
\bibliographystyle{apsrev4-1}
\newcommand{\be}{\begin{equation}}
\newcommand{\ee}{\end{equation}}
\newcommand{\bs}{\begin{split}}
\newcommand{\es}{\end{split}}
\newcommand{\R}[1]{\textcolor{red}{#1}}
\newcommand{\B}[1]{\textcolor{blue}{#1}}

\title{A Hybrid Rydberg Quantum Gate for Quantum Network}
\author{Yubao Liu}
\affiliation{Center for Gravitational Experiment, Hubei Key Laboratory of Gravitation and Quantum Physics, School of Physics, Huazhong University of Science and Technology, Wuhan, 430074, P. R. China}
\author{Lin Li}
\email{linli@hust.edu.cn}
\affiliation{Center for Gravitational Experiment, Hubei Key Laboratory of Gravitation and Quantum Physics, School of Physics, Huazhong University of Science and Technology, Wuhan, 430074, P. R. China}
\author{Yiqiu Ma}
\email{myqphy@hust.edu.cn}
\affiliation{Center for Gravitational Experiment, Hubei Key Laboratory of Gravitation and Quantum Physics, School of Physics, Huazhong University of Science and Technology, Wuhan, 430074, P. R. China}

\begin{abstract}
The high fidelity storage, distribution and processing of quantum information prefers qubits with different physical properties. Thus, hybrid quantum gates interfacing different types of qubits are essential for the realization of complex quantum network structures. A Rydberg-atom based physical quantum CZ gate is proposed to hybridly process the  polarisation-encoded single-photon optical qubit and the ``Schroedinger cat" microwave qubit.  The degradation of the fidelity under the influence of various noise channels, such as microwave cavity loss, sponetanous emission of atom states, and non-adiabaticity effect, etc, has been analyised through detailed theoretical analysis by deriving input-output relation of qubit fields. The feasibility and the challenges of the protocol within current technology are also discussed by analysing the possible experimental parameter settings. 
\end{abstract}

\maketitle
\noindent\emph{Introduction---}
A robust and functional quantum network\,\cite{pompili2021realization,daiss2021quantum} requires the simultaneous achievement of both high-fidelity local quantum operation and efficient distribution of quantum information between remote quantum nodes\,\cite{chou2007functional,PhysRevLett.78.3221}. The combination of these two capabilities would enable a number of important applications, such as long distance quantum communication\,\cite{duan2001long,stannigel2010optomechanical,muralidharan2016optimal,wallnofer2020machine,zhu2017experimental}, distributed quantum computation\,\cite{cohen2018deterministic,cuomo2020towards,van2016path,beals2013efficient,lim2005repeat} and quantum metrology\,\cite{zhou2020quantum,giovannetti2006quantum,giovannetti2011advances,toth2014quantum,chin2012quantum,joo2011quantum}. These two capabilities have been separately demonstrated in various experiments\,\cite{Kjaergaard2020sup,chen2021integrated}, but it remains very challenging to combine them into a single physical system. For example, microwave (MW) superconducting qubits feature high quantum operation fidelity\,\cite{arute2019quantum,Barends2014super} but suffer severe losses and decoherence during propagation\,\cite{Wendin_2017}. On the other hand, optical photons are ideal qubits for the purpose of long-distance quantum state distribution\,\cite{chen2021integrated}, but fault-tolerant quantum gates with optical photons remain elusive\,\cite{Slussarenko2019Photonic,RevModPhys.79.135}. Naturally, the realization of a practical quantum network would largely benefit from a hybrid platform\,\cite{PhysRevX.7.031001,PhysRevA.100.012307,PhysRevA.90.040502,PhysRevLett.108.063004,clerk2020hybrid,kurizki2015quantum,pritchard2014hybrid,schoelkopf2008wiring,PhysRevA.94.062301,RevModPhys.85.623,Wallquist_2009,morgan2020coupling} which bridges the gap between optical and MW qubits. Therefore, designing a physical photon-photon quantum gate that (1) hybridly processes qubits at different frequencies, (2) has high fidelity is important for achieving this target.

Quantum gates involving Rydberg atoms\,\cite{RevModPhys.82.2313} would be able to satisfy the above requirements. For realising a hybrid photon-photon quantum gate, highly excited Rydberg states are employed for interfacing with the MW qubits, while the hyperfine ground states and low lying excited states can be used to process the optical qubit. Meanwhile, the Rydberg states and hyperfine ground states has long lifetime\,\cite{beterov2009quasiclassical}, which could help keeping a high fidelity of the gate\,\cite{hao2015quantum}.

In this Letter, we propose a new scheme of such a Rydberg quantum CZ-gates, where the optical qubit is interfaced by eletromagnetic induced transparency (EIT) photon storage\,\cite{fleischhauer2002quantum,RevModPhys.75.457,gorshkov2007photon,wang2019efficient,tiarks2019photon} and the MW qubit is processed by high quality MW cavity QED system\,\cite{besse2018single,kono2018quantum,parkins2014microtoroidal,mabuchi2002cavity}. The system dynamics and the gate fidelity are analysed in details and the result shows that our protocol could provide a high fidelity hybrid quantum CZ gate within current technology\,\cite{reiserer2014quantum,tiarks2019photon}.

\begin{figure}[h]
\centering
\includegraphics[width=0.48\textwidth]{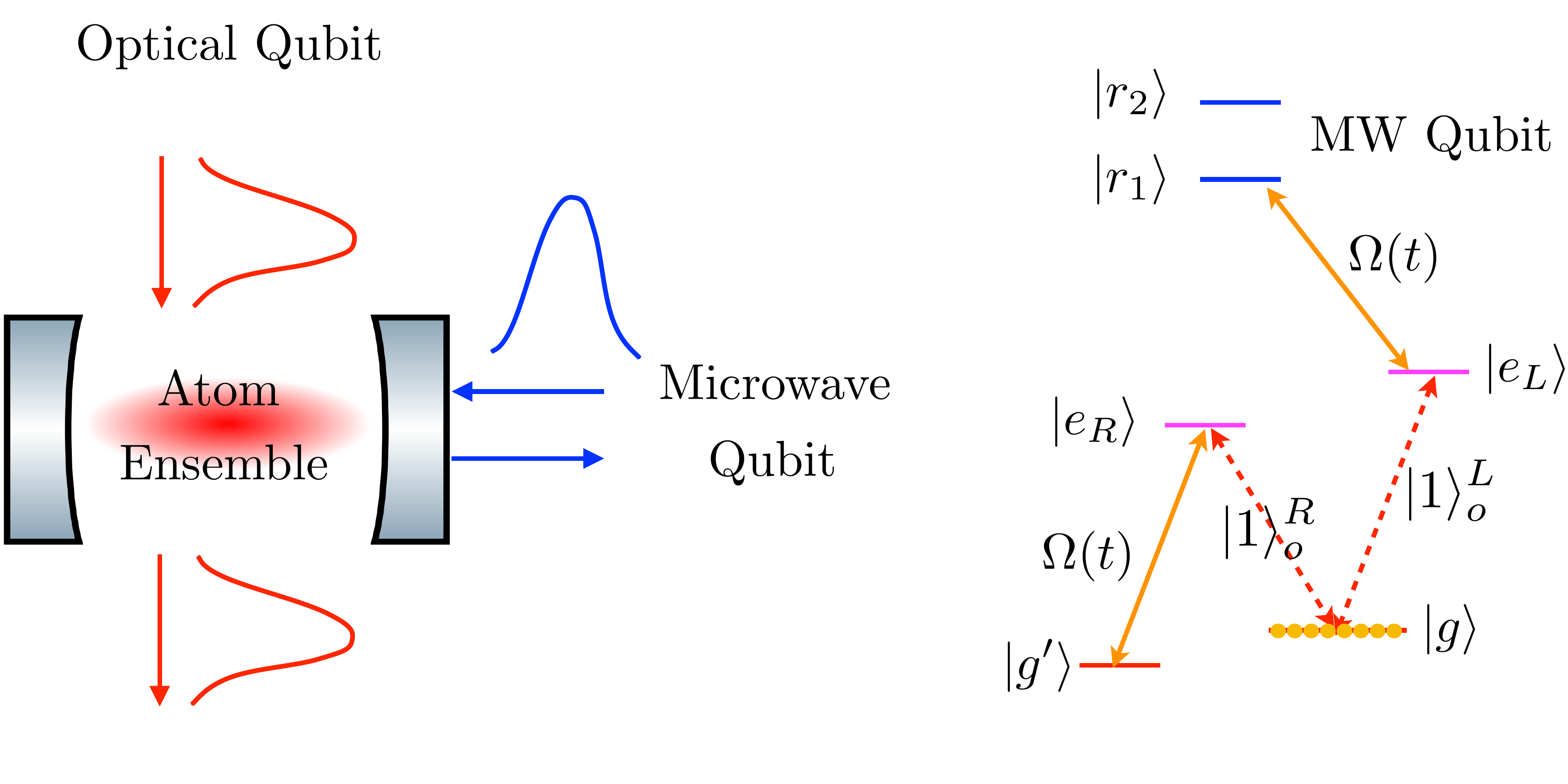}
\caption{Schematic setup. Left panel: an ensemble
 of Rb atoms trapped in a microwave\,(MW) cavity. Right panel: The relevant atomic levels used for the hybrid quantum CZ-gate. The atoms are initially populated on $|g\rangle$.}\label{fig:schematic_setup}
\end{figure}

\noindent{\it Design Concept---} Our model of a quantum hybrid CZ physical gate is schematically shown in Fig.\,\ref{fig:schematic_setup}.  The information of the optical qubit is encoded in the polarisations, while the information of the microwave qubit is encoded in the Schroedinger cat state. An ensemble of Rubidum (Rb) atoms is trapped in a MW resonator. The two optical qubit states interact with the gate via a $\Lambda-$type ($\left\vert{g}\right\rangle$-$\left\vert{e_R}\right\rangle$-$\left\vert{g'}\right\rangle$) and a ladder type system ($\left\vert{g}\right\rangle$-$\left\vert{e_L}\right\rangle$-$\left\vert{r_1}\right\rangle$),  where $(\left\vert{g}\right\rangle,\left\vert{g'}\right\rangle)$ are the hyperfine structures,  $(\left\vert{e_R}\right\rangle,\left\vert{e_L}\right\rangle)$ are the excited states. The $\Lambda$/ladder- EIT system can store single photon states with right/left circular polarisation ($|1\rangle^{R/L}_o$)\,\cite{tiarks2019photon,wang2019efficient}. The MW qubits is resonant with the two Rydberg states $\left\vert{r_1}\right\rangle$ and $\left\vert{r_2}\right\rangle$.

\begin{figure}[ht]
\centering
\includegraphics[width=0.45\textwidth]{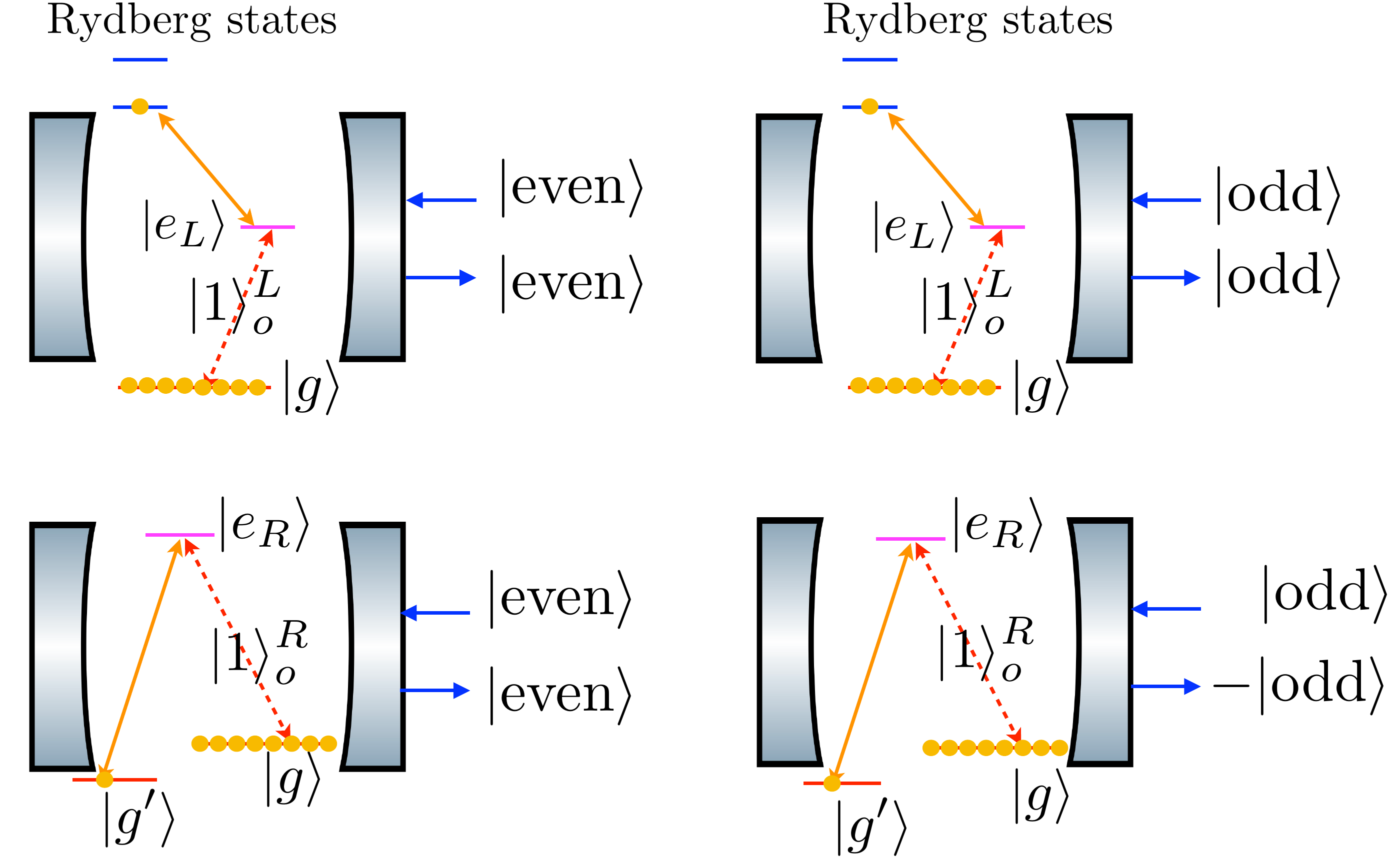}
\caption{Physical processes correspond to different qubit inputs of the quantum hybrid CZ-gate, with truth table in Tab.\,\ref{tab:truthtable}.}\label{fig:truthtable}
\end{figure}

During the EIT-storage, the MW qubit evolutes jointly with cavity field and the spin wave. For an input optical qubit $|1\rangle^R_o$, it will be stored by the $\Lambda$-EIT system in the form of atomic spin wave $|g\rangle-|g'\rangle$, and thus no atom will be excited to Rydberg states.  Therefore, the MW qubits will enter an empty cavity and the outgoing qubit fields simply get a $\pi-$phase shift. In particular, the input state $|\rm{even/odd}\rangle_m\propto|\alpha\rangle\pm|-\alpha\rangle$ will be transformed to $\pm(|\alpha\rangle\pm|-\alpha\rangle$).  However, for $|1\rangle^L_o$ input, the spin wave $|g\rangle-|r_1\rangle$ will be excited and the Rydberg level will be occupied. In this case, the MW qubit can strongly interact with the Rydberg atom once entering the MW resonator, therefore acquires no additional phase when it reflects back. These physical scenarios establish the CZ-truth table in Tab.\,\ref{tab:truthtable} and Fig.\,\ref{fig:truthtable}.\\

\begin{table}[h]
  \begin{center}
    \begin{tabular}{|c|c|c|c|}
    \hline
      Input state & Output state & Efficiency & Fidelity\\
      \hline
       $|1\rangle^R_o\otimes|{\rm even}\rangle_m$  & $+|1\rangle^R_o\otimes|{\rm even}\rangle_m$& $0.74$ & $0.923$ \\
      \hline
      $|1\rangle^R_o\otimes|{\rm odd}\rangle_m$ & $-|1\rangle^R_o\otimes|{\rm odd}\rangle_m$& $0.74$ &$0.923$\\
      \hline
      $|1\rangle^L_o\otimes|{\rm even}\rangle_m$ &$+|1\rangle^L_o\otimes|{\rm even}\rangle_m$&$0.45$&$0.969$\\
      \hline
      $|1\rangle^L_o\otimes|{\rm odd}\rangle_m$&$+|1\rangle^L_o\otimes|{\rm odd}\rangle_m$& $0.45$&$0.967$\\
       \hline
    \end{tabular}
    \caption{Truth table of the hybrid quantum CZ-gate, where the optical qubit is encoded in polarisations, and the microwave qubit is encoded in cat states: $|{\rm even}\rangle_m=|\alpha\rangle+|-\alpha\rangle$ and $|{\rm odd}\rangle_m=|\alpha\rangle-|-\alpha\rangle$. The inefficiency is due to the loss in the EIT storage process and the fidelity here is calculated considering the post-selection of the output optical single photon qubit (see Supplementary Material for details).  }\label{tab:truthtable}
  \end{center}
\end{table}

\noindent{\it Theoretical Model---}
We now setup a theoretical model for our scheme.
The Hamiltonian for the free-propagating electromagnetic (EM) field (e.g. along $+z$ direction) is given by:
\be
\hat H_{\rm EM}/\hbar=\frac{ic}{2L_a}\int_{-\infty}^{+\infty}\mathrm{d}z\left[\frac{\partial\hat{e}^{\dagger}}{\partial z}\hat{e}-\frac{\partial\hat{e}}{\partial z}\hat{e}^{\dagger}\right],
\ee
where the $\hat e^{(\dag)}$ is the annihilation(creation) operator of slowly varying amplitude of EM fields (for details , see Supplementary Material (SM).), $c$ and $L_a$ are the light speed in vacuum and the length of the atomic medium, respectively. In the following discussion, for the optical/MW qubit, $\hat e$ is written as $\hat \epsilon$/$\hat b$, respectively.

The atomic ensemble and the atom-light interaction are described by\,\cite{fleischhauer2002quantum,RevModPhys.75.457,gorshkov2007photon}:
\be
\begin{split}
&\hat H_{\rm atom}/\hbar=-\sum_{j=1}^{N}(\omega_{a}\hat\sigma_{aa}^{j}+\omega_{c}\hat\sigma_{cc}^{j}),\\
&\hat H_{\rm int}/\hbar=-\sum_{j=1}^{N}[\Omega(z_{j},t)e^{-i\upsilon_{\Omega}t+
ik_{\Omega}z_{j}}\hat\sigma_{ac}^{j}+g\hat{\varepsilon}\hat\sigma_{bc}^{j}]+{\rm h.c}.
\end{split}
\ee
Note that the atom described by this Hamiltonian has an energy structure $|b\rangle-|a\rangle-|c\rangle$ with the dark state distributed on $|b\rangle-|c\rangle$), represents both of the two possible EIT systems ($|g\rangle-|e_R\rangle-|g'\rangle$ and $|g\rangle-|e_L\rangle-|r_1\rangle$). The $\Omega(z_j,t)$ is the time-dependent control field and $g$ is the single-photon Rabi frequency.

The interaction between MW qubit and the cavity-QED (cQED) system is\cite{duan2004scalable,wang2005engineering}:
\begin{equation}
\begin{aligned}
\hat H_{\rm CQED}/\hbar = g_m\hat{a}_c\hat{\sigma}_{r_2 r_1}+\sqrt{\kappa}\hat{a}_c\hat{b}^{\dag}(0)+{\rm h.c,}
\end{aligned}
\end{equation}
where the first term describes the interaction between MW cavity mode $\hat a_c$ and the two Rydberg states $|r_1\rangle, |r_2\rangle$ with single-photon Rabi frequency $g_m$, and the second term represents the interaction between free propagating MW qubits and the MW cavity mode with strength $\sqrt{\kappa}$\,\cite{lin2006one}.

The atomic states and the cavity field will also be affected by the noisy environment, such as the vacuum field which induces spontaneous emission, the collision of atoms and the cavity loss etc. We denote these effects by $\gamma_{\mu\nu}$ which is the decoherence rate between $|\mu\rangle$ to $|\nu\rangle$ ($\mu,\nu$ are among $|g'\rangle,|g\rangle,|e_R\rangle, |e_L\rangle$), $\gamma_s$ as the decay rate of Rydberg state and $\kappa_s$ to be the MW cavity loss rate (note that together with $\kappa$ it contributes to the cavity bandwidth). These factors will bring decoherence to the quantum state and affect the gate fidelity.

\noindent{\it Possible Physical Realisations---} For later analysis and simulation, we consider the following possible physical realisation. The inital Rb atomic state is prepared in $|g\rangle=|5S_{1/2},F=m_F=2\rangle$, with its neighbouring ground state is $|g'\rangle=|5S_{1/2},F=m_F=1\rangle$, and the intermediate quantum states are $|e_R\rangle=|5P_{3/2},F=2,m_F=1\rangle$ and $|e_L\rangle=|5P_{3/2},F=m_F=3\rangle$. Besides, the Rydberg states are chosen to be $|r_1\rangle=|69S_{1/2},F=m_F=2\rangle$ and $|r_2\rangle=|69P,F=3,m_F=3\rangle$.

For the EIT storage of optical qubit: (1) The atomic ensemble has length $L_a=0.4$\,mm with density $\rho\sim 8\times 10^{11}$\,cm$^{-3}$ (i.e. roughly $6\times 10^4$ atoms) and interaction cross-sectional diameter $w=8\,\mu$m. The single-photon Rabi frequency for the $|e_R\rangle-|g\rangle$ and $|e_L\rangle-|g\rangle$ transitions can be derived as $g_L/2\pi=29$\,kHz and $g_R/2\pi=12$\,kHz, respectively. (2) The control field Rabi frequency contributed of the EIT processes is $\Omega(t)=\Omega_0(2+\tanh[20(t-18\mu s]-\tanh[20(t-2\mu s)])/2$, where $\Omega_0=2\pi\times 30$\,MHz and the storage time is $16\,\mu$s.  (3) The decay rates of atom states are $\gamma_{eg}/2\pi=3$\,MHz, $\gamma_{r_1g}/2\pi=3.5$\,kHz\,\cite{schmidt2020dark} and $\gamma_{g'g}/2\pi=16$\,Hz\,\cite{PhysRevA.87.031801}. This small $\gamma_{r_1g}$ requires the temperature of the atomic ensemble below $\sim 0.2\,\mu$K.

For the MW cQED system: the MW cavity has a quality factor $\sim 10^4$, the Rydberg state $|r_2\rangle$ has decay rate $\gamma_s/2\pi\sim 4.78$\,kHz\,\cite{beterov2009quasiclassical}, and the single-photon Rabi frequency $g_m/2\pi=2.72$\,MHz. Such a strong $g_m$ demands a large dipole moment and a small interaction volume, which indicates a possible realisation by combining  Rydberg atoms with a superconducting circuit\,\cite{petrosyan2009reversible,pritchard2014hybrid,morgan2020coupling}.


\noindent{\it Field operator evolution---}The above Hamiltonians can be solved using the standard method under the adiabatic conditions. Namely, the coherent control field will be switched on and off slowly enough to avoid the transition from dark state to bright state, and the qubit field changes negligibly during the switching of the control field. The resultant evolution of the optical qubit is given by (see SM for detailed derivations):

\begin{equation}\label{eq:epsilon}
\begin{aligned}
\left(\frac{\partial}{\partial t}+v\frac{\partial}{\partial z}\right)\hat{\varepsilon}(z,t)=&A\hat{\varepsilon}(z,t)+{c^2(1-\eta)^2C}\frac{\partial^2}{\partial z^2}\hat{\varepsilon}(z,t)\\&+{c^3(1-\eta)^3D}\frac{\partial^3}{\partial z^3}\hat{\varepsilon}(z,t),
\end{aligned}
\end{equation}
where the $A>0$$/<0$ represents the gain/attenuation of $\hat\epsilon$, corresponds to the retrieving/writing of the optical qubit, respectively. Coefficients $C$ and $D$ describe the qubit pulse spreading due to the limited EIT window and the higher-order distortion of the qubit waveform. These coefficients can be written as:
\be
\begin{split}
&A=\eta\left[\frac{1}{\Omega}\frac{\partial\Omega}{\partial t}-\gamma_{bc}\right],\quad D=\frac{\eta}{|\Omega|^2},\\
&C=\frac{\eta}{|\Omega|^2}\left[(2\gamma_{bc}+\gamma_{ba})-2\left(\frac{1}{\Omega^*}\frac{\partial\Omega^*}{\partial t}+\frac{2}{\Omega}\frac{\partial\Omega}{\partial t}\right)\right],
\end{split}
\ee
in which $\eta=g^2N/(g^2N+|\Omega|^2)$ with $N$ the total number of atoms.

Solving the above equation leads to the evolution of the field in the $k$-domain:
\begin{equation}\label{eq:opticalevolution}
\begin{aligned}
&\hat{\varepsilon}(k,T)=C^{oi}_{1}(k,T)\hat{\varepsilon}(k,0)+C^{oi}_{2}(k,T)\hat n(k),\\
&{\rm with}\quad
C^{oi}_{1}(k)={\rm exp}\left\{-i\int_{0}^{T}\mathrm{d}t[kv-k^3c^3D(1-\eta)^3]\right\}\\
&\qquad\qquad\qquad\quad{\rm exp}\left\{\int_{0}^{T}\mathrm{d}t\left[A-{k^2c^2(1-\eta)^2}C\right]\right\},
\end{aligned}
\ee
and the optical qubit takes time $T$ to finish its full writing and retrieving process, $i$ represent left/right polarsiations. The second exponential factor describes gain/attentuation of $\hat\epsilon$. The above formula also captures various loss channels, for example, the loss due to non-adiabaticity (non-zero $d\Omega/\Omega dt$) and the dissipation of various energy levels ($\gamma_{bc},\gamma_{ba}$). As a bosonic operator, $\hat \epsilon$ should satisfy Bosonic commutation relation, thereby other degrees of freedom should be added onto the above formula to conserve the phase space\footnote{This is the direct result of \emph{fluctuation-dissipation} theorem, which has been extensively discussed by Caves\,\cite{Caves1982} for optical systems.}. For example, 
for the output optical field from the atom ensemble, we should add to $\hat \epsilon$ those noise fields $\hat n$ which are associated with various loss channels and $|C^{oi}_{1}|^2+|C^{oi}_2|^2=1$ according to\,\cite{Caves1982}.

The evolution of the MW field through cQED process is (for a complete derivation, see SM):
\be
\hat b_{\rm out}(\omega)=C_1(\omega)\hat b_{\rm in}(\omega)+C_2(\omega)\hat n_{\rm in}(\omega),
\ee
where $\hat n_{\rm in}$ represents all possible noise contributions (includes the cavity loss and Rydberg state decay) and 
\be
\begin{split}
&C_1(\omega)=\frac{i\omega+g_m^2/(i\omega-\gamma_s)-(\kappa_s-\kappa)/2}
{i\omega+g_m^2/(i\omega-\gamma_s)-(\kappa_s+\kappa)/2},\\
&C_2(\omega)=\frac{i\omega-\gamma_{\rm eff}}{i\omega-\gamma_s}\frac{\sqrt{\kappa\kappa_s}}{i\omega+g_m^2/(i\omega-\gamma_s)-(\kappa_s+\kappa)/2},\\
&|C_1(\omega)|^2+|C_2(\omega)|^2=1,
\end{split}
\ee
with $\gamma_{\rm eff}=\sqrt{\gamma_s^2+2g_m^2\gamma_s/\kappa_s}$.
In the strong coupling region, where $g_m^2\gg \gamma \kappa$, the output field has a zero phase shift compared to the input field. However, in case the Rydberg state is empty when the optical qubit is $|1\rangle_R$, we have $g_m=0$ and the input-output relation reduces to the form of a simple cavity, and the phase shift $\approx \pi$. In deriving the above formula, we have used the linear approximation $\langle\sigma_z\rangle\approx -1/2$, under which the cQED dynamics becomes linear. This linear dynamics is essential for a high fidelity quantum gate. This approximation valids in the strong coupling region and the intra-cavity quanta is small (for detailed discussions, see SM).

\noindent{\it Loss and mode shape distortion---}
The loss and mode shape distortion can affect the gate fidelity. Here we give an analysis of these effects by studying the quantum state evolution.

A single photon quantum state can be described as: $|\Psi(t)\rangle=\int dzf(z,t)\hat \epsilon^\dag(z)|\rm vac\rangle=\int dkf(k,t)\hat \epsilon^\dag(k)|\rm vac\rangle$, where $f(z,t)$ is the normalised state waveform, with its Fourier conjugate $f(k,t)$. The concrete form of $f(k,t)$ can be determined by Eq.\,\eqref{eq:opticalevolution}. The waveform evolution of the optical qubit during writing and retrieving process is simulated using the parameters listed in the previous sections with the results shown in Fig.\,\ref{fig:opticalevolution}.
\begin{figure}[ht]
\centering
\includegraphics[width=0.5\textwidth]{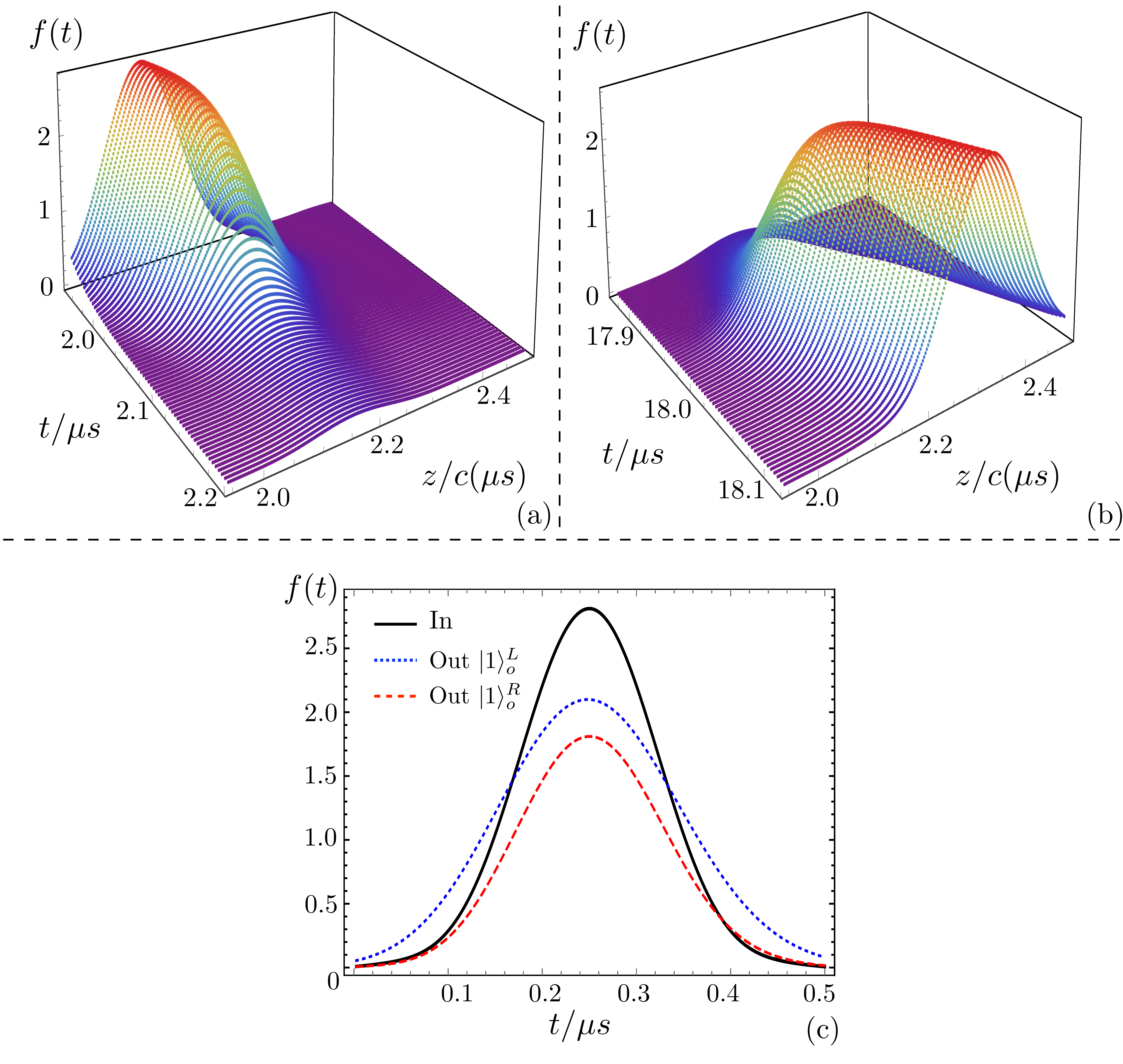}
\caption{Optical qubit waveform evolution: The upper panel describes the writing of optical qubit into the atom ensemble (a) and the retrieval process after the $15\,\mu$s storage time (b). The lower panel shows the \emph{un-normalised} output qubit waveform ($|1\rangle^L_o$ as blue dotted and $|1\rangle^R_o$ as red dashed) retrived from atom ensemble, compared with the input pulse (black solid). These waveform needs to be normalised for \emph{post selection scheme} (c). }\label{fig:opticalevolution}
\end{figure}
The effect of the spin wave loss to the optical qubit can be phenomenologically written as $|\Psi\rangle_{\rm out}^i=C^{oi}_1|1\rangle^i_{q\rm out}\otimes|0\rangle_n^i+C^{oi}_2|0\rangle^i_{q\rm out}\otimes|1\rangle_n^i$, where the $i=L/R$ represent the left/right polarisations and the $C^{oi}_{1,2}$ are given by Eq.(\R{6}). Such a qubit-noise entanglement leads to decoherence. Since the photonic qubit is a single-photon state, therefore \emph{post selection scheme} can be implemented. Generally speaking, qubits with different polarisations could suffer different losses, that is, $C^{oL}_1\neq C^{oR}_2$. For an input state $|\Psi\rangle=(|1\rangle^{L}_{\rm in}+|1\rangle^{R}_{\rm in})/\sqrt{2}$, the output state after the post-selection with $C^{oL}_1\neq C^{oR}_2$ is apparently deviated from the ideal case. Therefore, to obtain a high gate fidelity after the post-selection, the method of balancing the loss of the two different polarisations can be applied\,\cite{daiss2021quantum}, at the cost of decreasing the post-selection probability\,\footnote{Actually, the distortion of the output state due to unbalanced $C^{o L/R}_{1,2}$ can be solved by adding a unitary rotation. However, such a rotation will be different for different input states while the tuning the loss to balance $C^{o L/R}_{1,2}$ is robust to different input states.}.

For a MW cat-state qubit $|\alpha_{\rm in/out}\rangle\pm|-\alpha_{\rm in/out}\rangle$, with
\be
|\alpha_{\rm in/out}(t)\rangle\propto{\rm exp}\left[\alpha_{0/1}\int dx f_{\rm in/out}(x,t)\hat b_{\rm in}^\dag(x)\right]|\rm vac\rangle,
\ee
where $\alpha_{0/1}$ is the input/output coherent amplitude and $f_{\rm in/out}(x,t)$ is the corresponding mode shape.  Different from the single-photon state, the post-selection scheme can not be used for cat state. The response, the MW cavity loss and the finite life time of the Rydberg states will affect the qubit mode shape, of which the latter two factors will also introduce qubit decoherence.

 We define the mode shape distortion as $1-\Lambda=1-\int dx f_{\rm in}^*(x)f_{\rm out}(x)$. For an occupied Rydberg state and in the strong coupling region, we have $\hat b_{\rm out}\approx-\hat b_{\rm in}$ thereby the mode shape suffers negligible distortion, i.e. $1-\Lambda\approx 1$. On the contrary, for an unoccupied Rydberg state, an empty cavity with smaller bandwidth has stronger response to the input qubit. This is  why  $1-\Lambda$ decreases as the increase of $\kappa$ in Fig.\,\ref{fig:cqedwaveform}. Moreover, the MW cavity loss can also affect the coherent amplitude $\alpha$.

\begin{figure}[h]
\centering
\includegraphics[width=0.4\textwidth]{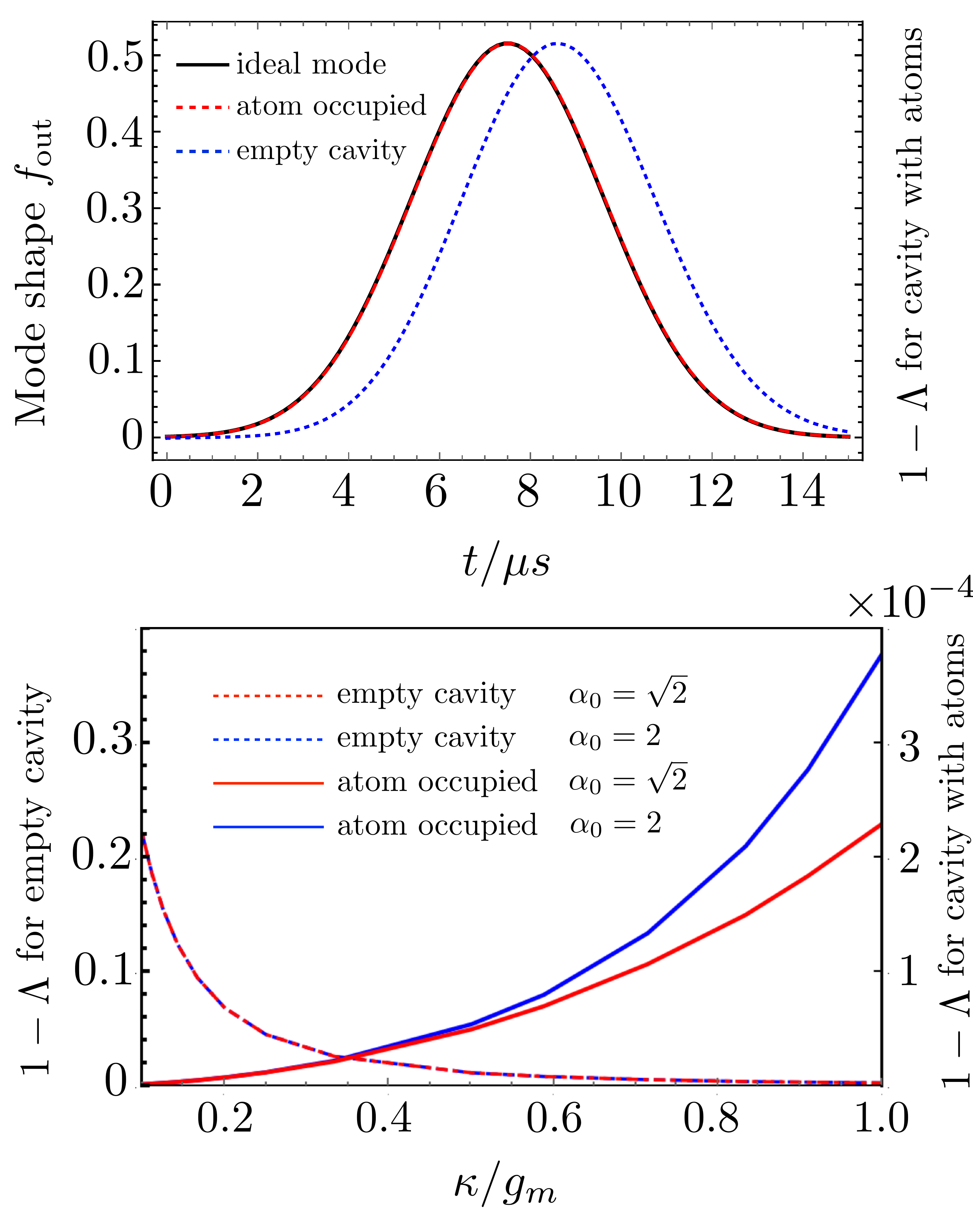}
\caption{Upper panel: the mode shape of output MW qubits when $\kappa/g_m=0.2$ and the cavity is lossless. Lower panel: the effect of coupling rate $\kappa$ to the phase variation and mode shape error between input and output of a lossless cavity.  The values of $g_m/2\pi$, $\gamma_s/2\pi$ are set to be $2.72$\,MHz, $4.78$\,kHz.}\label{fig:cqedwaveform}
\end{figure}

\begin{figure}[ht]
\centering
\includegraphics[width=0.5\textwidth]{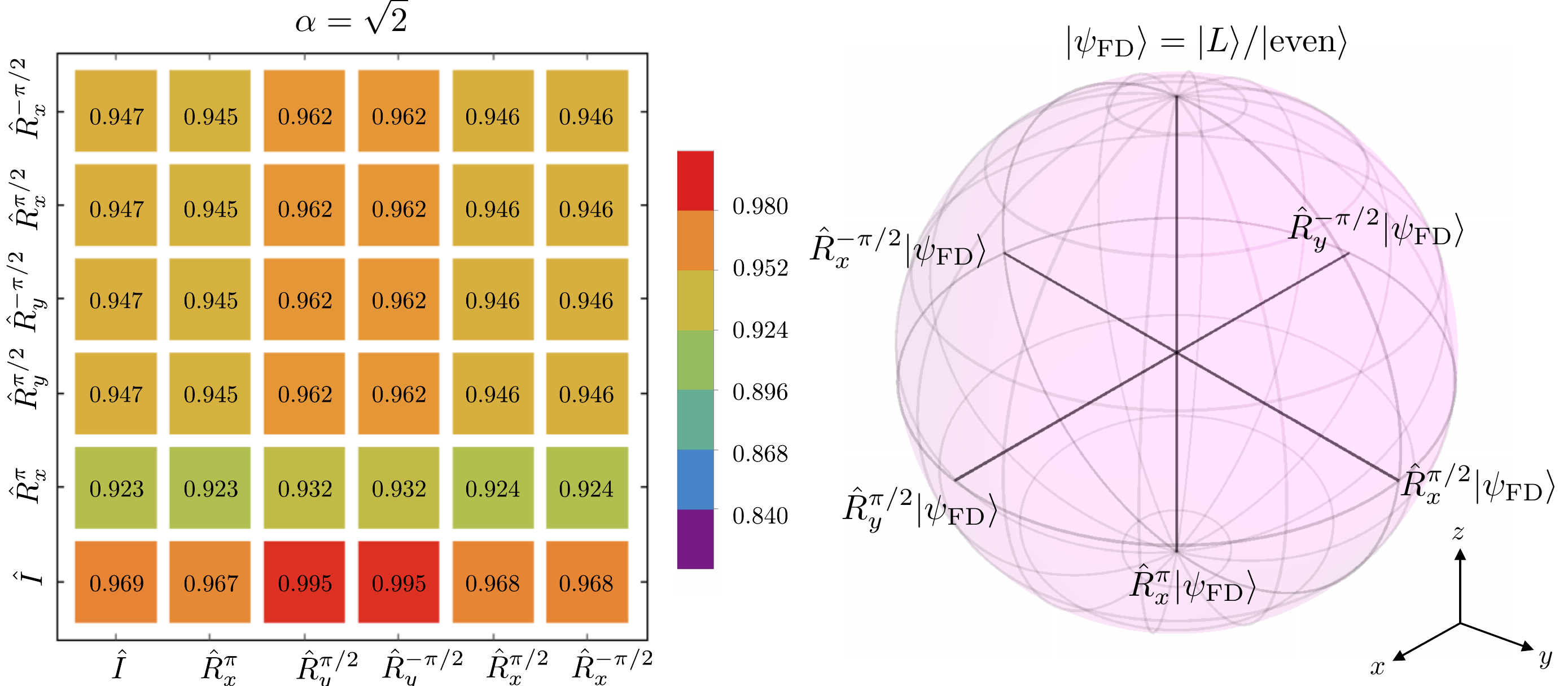}
\caption{The fidelity of the hybrid quantum gate with input product states when $\kappa_s/\kappa=10^{-3}$. The horizontal and vertical axis represent optical and MW qubit states, respectively. The color represents the fidelity values.  For illustrative purpose, we define a fiducial state $|\psi\rangle_{\rm FD}=|1\rangle^L_o/|\rm even\rangle$, and other states are represented rotations $\hat R_j^\theta$ on the Bloch sphere where $j$ is the axis index and $\theta$ is the rotation angle.}\label{fig:fidelitymain}
\end{figure}

\noindent{\it Fidelity---} General method to calculate the fidelity of the qubit entangled with the environmental noise is performed by $\mathcal{F}=\langle \Psi_{\rm ideal}|{\rm Tr}_E[\hat \rho_{\rm out}]|\Psi_{\rm ideal}\rangle$\,\cite{jozsa1994fidelity}, in which we trace out the noise channel in the output density matrix. We leave the details of fidelity calculation in the SM. Fig.\,\ref{fig:fidelitymain} shows the fidelities (considering \emph{post-selection scheme} for optical qubits) for those processes with separable MW and optical input states. To demonstrate the quantum nature of our CZ gate, we choose the product input state which are represented by rotations in the MW/optical bloch sphere $|\psi\rangle_{\rm in}=|\psi\rangle_q^o\otimes|\psi\rangle_q^{\rm MW}=\hat R^\theta_i|1\rangle^L_o\otimes\hat R^{\theta'}_j|\rm even\rangle$ and the gate operations create entanglement between optical and MW state. In computing the fidelity, we choose the MW cavity loss to be $\kappa_s/\kappa=0.001$\,\cite{Cleland2012} and $\alpha=\sqrt{2}$.

The fidelity after the post-selection is affected by the following factors. (1) The optical (normalised) waveform distortion after propagating through the EIT medium; (2) the MW cavity loss which introduces the decoherence, and the response of the MW cavity, which induces the dispersive distortion of the MW mode shape (see Fig.\,\ref{fig:cqedwaveform}); (3) the distortion of the MW mode shape due to the small nonlinearity of the cQED system. We have plotted the effect of the latter two factors in Fig.\,\ref{fig:average_fidelity}, in which the average fidelity $\bar{\mathcal{F}}$ is obtained by averaging the fidelity over all separable input states shown in Fig.\,\ref{fig:fidelitymain}.

\begin{figure}[h]
\centering
\includegraphics[width=0.42\textwidth]{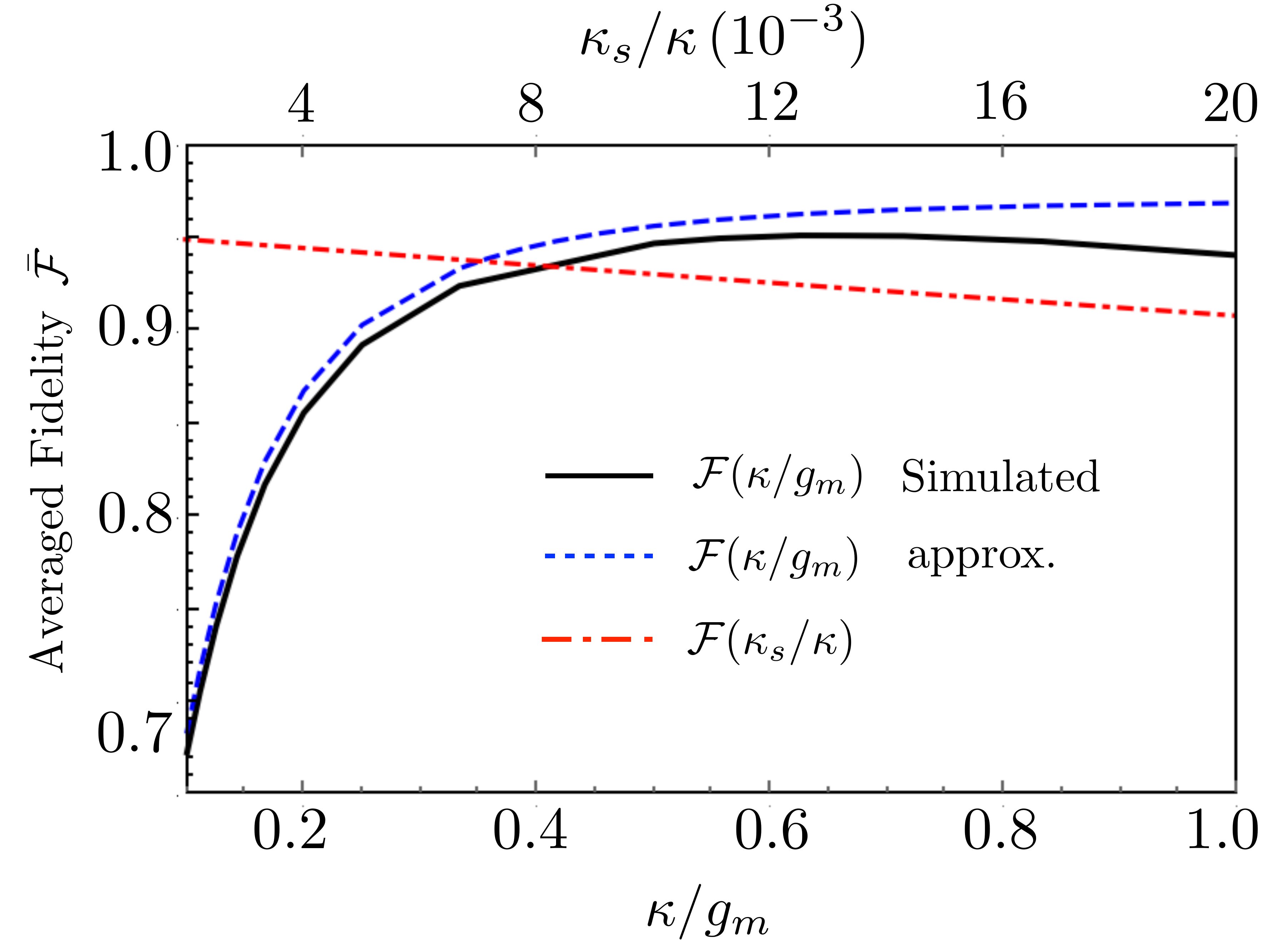}
\caption{The averaged fidelity of the hybrid quantum CZ gate shown as a function of cavity loss $\kappa_s$ and cavity bandwidth $\kappa$, when $g_m/{2\pi}=2.72$\,MHz, $\kappa/{2\pi}=2.0$\,MHz and $\gamma_s/{2\pi}=4.78$\,kHz. The solid line and the dashed line are the result obtained by full cQED simulation and linear approximation $\langle \hat \sigma_z\rangle=-1/2$.}\label{fig:average_fidelity}
\end{figure}

\noindent{\it Conclusion---}
In summary, we have proposed a scheme to realise an optical-microwave hybrid quantum CZ gate by ultising the properties of Rydberg atom states, which could be a building block for the future distributed quantum computing and network. The interaction between qubit fields and the atom systems are thoroughly analysed. The performance of the proposed gate under the influence of various noise channels is studied and we conclude that a high fidelity quantum gate can be in principle achieved. From our analysis on the parameter settings (for details, see SM), the main experimental challenge for realising a high fidelity gate is (i) The low decoherence rate of Rydberg atom states $\gamma_{r_1 g}$ requires either spatial confinement of Rydberg atoms, or employ ultra-code atoms; (ii) design and integration of high-Q microwave resonator compatible with Rydberg atoms to reach the strong coupling of cavity mode and atom state. Although this work focus on analysing the main conceptual features of this hybrid quantum gate scheme, we have used the current state of art parameters to simulate the gate performance. Our result indicates that, with the development of the technology, such a scheme of hybrid quantum gate could find applications in future development of remote quantum gate, long distance quantum teleportation and quantum network establishment.

\noindent{\it Acknowledgement---} L.L. and Y.M. are supported by the university start-up fundings provided by Huazhong University of Science and Technology. The authors thank Daiqin Su, Saijun Wu, Shengjun Yang, Kuan Zhang, Shuai Shi, Zhiyuan Wei, Yafen Cai, Zebing Zhou and Yanbei Chen for helpful discussions. Y.M. thanks Yuelong Liu for his administrative support. L.L. is supported by the National Natural Science Foundation of China (Grant No. 12004127).

\bibliography{reference}

\appendix
\section{Hamiltonian for a 1-D free-propagating EM fields}
The Hamiltonian describes an one-dimensional freely propagating quantum eletromagnetic (EM) field is derived as follows. In the frequency domain, the general form of the free EM Hamiltonian is:
\begin{equation}
H_{f}=\hbar c\int_{0}^{\infty}\mathrm{d}k |k|\hat{a}_{k}^{\dagger}\hat{a}_{k}.
\end{equation}
For uni-directional field pulse with center spatial frequency $k_0$, it can be written as:
\begin{equation}
H_{f}=\hbar c\int_{-k_0}^{\infty}\mathrm{d}k \left(k_0+k\right)\hat{a}_{k_0+k}^{\dagger}\hat{a}_{k_0+k},
\end{equation}
Since we are interested in the case $k\ll k_{0}$, we can approximate the above formula as:
\begin{equation}\label{eq:Hffrequency}
H_{f}=\hbar c\int_{-\infty}^{+\infty}\mathrm{d}k (k_0+k)\hat{a}_{k_0+k}^{\dagger}\hat{a}_{k_0+k},
\end{equation}
Now we define the Fourier transform of the field operator in $z$ space:
\begin{equation}
\hat{a}_{z}=\int_{-\infty}^{+\infty}\hat{a}_{k_0+k}e^{-ikz}\mathrm{d}k,\quad \hat{a}_{k_0+k}=\int_{-\infty}^{+\infty}\hat{a}_{z}e^{ikz}\mathrm{d}z.
\ee
Substituting them into Eq.\,\eqref{eq:Hffrequency}, and choosing rotating frame $e^{-ik_0z}$ we obtain:
\begin{equation}
H_{f}=\frac{i\hbar c}{2}\int_{-\infty}^{+\infty}\mathrm{d}z\left[\frac{\partial\hat{a}_{z}^{\dagger}}{\partial z}\hat{a}_{z}-\frac{\partial\hat{a}_{z}}{\partial z}\hat{a}_{z}^{\dagger}\right],
\end{equation}
If the quantisation volume has length $L$, the quantum optical field operator can be redefined as $\hat{\varepsilon}=\sqrt{L}\hat{a}_{z}$, which leads to:
\begin{equation}
 H_{f}=\frac{i\hbar c}{2L}\int_{-\infty}^{+\infty}\mathrm{d}z\left[\frac{\partial\hat{\varepsilon}^{\dagger}}{\partial z}\hat{\varepsilon}-\frac{\partial\hat{\varepsilon}}{\partial z}\hat{\varepsilon}^{\dagger}\right],
\end{equation}
which is the Hamiltonian we used in the main text.

\section{Derivation of the optical qubit evolution}
In the model discussed in the main text, the weak probe field $\hat{\epsilon}$ resonantly couples to $\left\vert{b}\right\rangle-\left\vert{a}\right\rangle$, and a coherent control field with Rabi frequency $\Omega(z,t)$ couples to the excited level $\left\vert{a}\right\rangle-\left\vert{c}\right\rangle$, and the Hamiltonian in the interaction picture is\,\footnote{The material here is a re-derivation of the classic Fleishhauer-Lukin paper\,\cite{fleischhauer2002quantum} using our own approach}:
\begin{equation}
\begin{aligned}
H=&\frac{i\hbar c}{2L}\int_{-\infty}^{+\infty}\mathrm{d}z\left[\frac{\partial\hat{\varepsilon}^{\dagger}}{\partial z}\hat{\varepsilon}-\frac{\partial\hat{\varepsilon}}{\partial z}\hat{\varepsilon}^{\dagger}\right]\\
&-\hbar\sum_{j=1}^{N}[\Omega(z_{j},t)\hat\sigma_{ac}^{j}+g\hat{\varepsilon}\hat\sigma_{bc}^{j}+{\rm h.c}],
\end{aligned}
\end{equation}
where $g$ is the atom-field coupling constant, $\hat\sigma_{\mu\upsilon}=\left\vert{\mu}\right\rangle\left\langle{\upsilon}\right\vert$ is the transition operator of atom from $\left\vert{\upsilon}\right\rangle$ to $\left\vert{\mu}\right\rangle$.  

To write it in a continuous form, we can introduce the following operator:
\begin{equation}
\begin{aligned}
\hat{\sigma}_{\mu\upsilon}(z,t)=\frac{1}{n(z)}\sum_{j=1}^{n(z)}\hat{\sigma}_{\mu\upsilon}^{j}(z,t),
\end{aligned}
\end{equation}
where $n(z)\gg1$ is the sum number of atoms at position $z$, and $\sum_{j=1}^{N}$ can be replaced by with $\int\mathrm{d}zn(z)$ in the continuum limit. Using these continuous atom-field operators, the Hamiltonian can be written as:
\begin{equation}
\begin{aligned}
\hat{H}&=\frac{i\hbar c}{2L}\int_{-\infty}^{+\infty}\mathrm{d}z\left[\frac{\partial\hat{\varepsilon}^{\dagger}}{\partial z}\hat{\varepsilon}-\frac{\partial\hat{\varepsilon}}{\partial z}\hat{\varepsilon}^{\dagger}\right]-\int_{0}^{L}\mathrm{d}zn(z)\\&\cdot[\hbar g\hat{\sigma}_{ab}(z,t)\hat{\varepsilon}(z,t)+\hbar\Omega(z,t)\hat{\sigma}_{ac}(z,t)+{\rm h.c.}],
\end{aligned}
\end{equation}
where we assume that the atoms ensemble has the same length as $L$, $N$ is the total number of atoms and $n(z)$ is the linear atom number density in propagation direction. The communication relations of the atomic and field operators are:
\begin{equation}
\begin{aligned}
&\left[\hat{\sigma}_{\mu\upsilon}(z,t),\hat\sigma_{\alpha\beta}(z',t)\right]\\&\quad=\frac{1}{n(z)}[\hat\sigma_{\mu\beta}(z,t)\delta_{\upsilon\alpha}-\hat\sigma_{\upsilon\alpha}(z,t)
\delta_{\mu\beta}]\delta(z-z'),\\
&\left[\hat{\epsilon}(z,t),\hat{\epsilon}^{\dagger}\right(z',t)]=L\delta(z-z').
\end{aligned}
\end{equation}
The corresponding Heisenberg-Langevin equations are:
\begin{equation}
\begin{split}
&\dot{\hat{\sigma}}_{ba}=-\gamma_{ba}\hat{\sigma}_{ba}+ig\hat{\varepsilon}(\hat{\sigma}_{bb}-\hat{\sigma}_{aa})+i\Omega\hat{\sigma}_{bc}+\hat F_{ba},\\
&\dot{\hat{\sigma}}_{bc}=-\gamma_{bc}\hat{\sigma}_{bc}+i\Omega^{*}\hat{\sigma}_{ba}-ig\hat{\varepsilon}\hat{\sigma}_{ac}+\hat F_{ac},
\end{split}
\end{equation}
where $\gamma_{\mu\upsilon}$ is the atom decay rate from level $\upsilon$ to $\mu$, and $\hat F_{\mu\upsilon}$ is corresponding Langevin noise operators.

The optical qubit propagation equation is as (where we assume the atomic medium is homogeneous, i.e. $n(z)=\rm{const}.$ and $n(z)L=N$):
\begin{equation}
\left(\frac{\partial}{\partial t}+c\frac{\partial}{\partial z}\right)\hat{\epsilon}(z,t)=igN\hat{\sigma}_{ba}(z,t),
\end{equation}

For the proposal discussed in the main text, the initial state the first-order perturbation of the equations of motion is:
\begin{equation}
\begin{split}
&\dot{\hat{\sigma}}_{ba}=-\gamma_{ba}\hat{\sigma}_{ba}+ig\hat{\epsilon}+i\Omega\hat{\sigma}_{bc}+\hat F_{ba},\\
&\dot{\hat{\sigma}}_{bc}=-\gamma_{bc}\hat{\sigma}_{bc}+i\Omega^{*}\hat{\sigma}_{ba}+\hat F_{ac}.
\end{split}
\end{equation}
Since the Raman process effectively only involve the two hyperfine structure levels, therefore we eliminate the excited level and obtain:
\begin{equation}\label{eq:sigmabaeliminated}
\begin{split}
&\left(\frac{\partial}{\partial t}+c\frac{\partial}{\partial z}\right)\tilde{\varepsilon}(z,t)=\frac{gN}{\Omega^{*}}\left(\frac{\partial}{\partial t}+\gamma_{bc}\right)\tilde{\sigma}_{bc}-\frac{gN}{\Omega^{*}}\hat{F}_{ac},\\
&\tilde{\sigma}_{bc}=-\frac{g\tilde{\varepsilon}}{\Omega}-\frac{i}{\Omega}\left(\frac{\partial}{\partial t}+\gamma_{ba}\right)\left[-\frac{i}{\Omega^{*}}\left(\frac{\partial}{\partial t}+\gamma_{bc}\right)\tilde{\sigma}_{bc}\right]\\&\quad\quad+\frac{1}{\Omega}\left(\frac{\partial}{\partial t}+\gamma_{ba}\right)\frac{F_{bc}}{\Omega^{*}}+\frac{iF_{ba}}{\Omega},
\end{split}
\end{equation}

Now, we perform the adiabatic approximation to these equations. Basically, the coherent control field is switch on and off slowly enough to avoid the transition from dark state to bright state. This means the terms related to time evolution of Eq.\,\eqref{eq:sigmabaeliminated} can be ignored, and the adiabatic condition can be expressed as:
\begin{equation}
\left|\frac{1}{\Omega}\frac{\partial \Omega}{\partial t}\right|\ll\frac{|\Omega|^{2}}{\gamma_{ba}},
\left|\frac{1}{\Omega}\frac{\partial\Omega}{\partial t}\right|\ll|\Omega|,
\left|\frac{1}{\Omega}\frac{\partial^2\Omega}{\partial t^2}\right|\ll|\Omega|^2,
\end{equation}
which leads to:
\begin{equation}
\left(\frac{\partial}{\partial t}+c\frac{\partial}{\partial z}\right)\hat{\varepsilon}(z,t)=\frac{gN}{\Omega^{*}}\frac{\partial \hat{\sigma}_{bc}}{\partial t},\quad
\hat{\sigma}_{bc}=-\frac{g\hat{\varepsilon}}{\Omega}.
\end{equation}
Further eliminating the $\hat\sigma_{bc}$ gives:
\begin{equation}
\left(\frac{\partial}{\partial t}+c\frac{\partial}{\partial z}\right)\hat{\varepsilon}(z,t)=\frac{g^{2}N}{|\Omega|^{2}\Omega}\frac{\partial\Omega}{\partial t}\hat{\varepsilon}-\frac{g^{2}N}{|\Omega|^{2}}\frac{\partial\hat\varepsilon}{\partial t},
\end{equation}
which can be written as:
\begin{equation}
\left(\frac{\partial}{\partial t}+v\frac{\partial}{\partial z}\right)\tilde{\varepsilon}(z,t)=\beta\tilde{\varepsilon}(z,t),
\end{equation}
Here, we define refractive index $n_g=1/(1-\eta)$ with $\eta=g^2N/(g^2N+|\Omega|^2)$, the group velocity $v$ and the decay rate $\beta$ are:
\be
v=c/n_g,\quad\beta=\frac{\eta}{\Omega}\frac{\partial\Omega}{\partial t}.
\ee
These equations describe an ideal EIT storage. When we reduce the control light, the speed of $\tilde{\epsilon}$ is slowed down, which is the EIT slow-light effect. The writing and retrieving processes correspond to $\beta<0$ and $\beta>0$, where the control field is switched off and on adiabatically, respectively.

Next we analyse the imperfections of the EIT storage, e.g. the non-adiabaticity and finite $\gamma_{bc},\gamma_{ba}$. We approximate the $\tilde{\sigma}_{bc}$ as $g\hat\varepsilon/\Omega$, leads to:
\begin{equation}
\begin{aligned}
\tilde{\sigma}_{bc}&=-\frac{g\hat{\varepsilon}}{\Omega}-\frac{i}{\Omega}\left(\frac{\partial}{\partial t}+\gamma_{ba}\right)\left[-\frac{i}{\Omega^{*}}\left(\frac{\partial}{\partial t}+\gamma_{bc}\right)\frac{g\hat\varepsilon}{\Omega}\right]\\&+\frac{1}{\Omega}\left(\frac{\partial}{\partial t}+\gamma_{ba}\right)\frac{\hat F_{bc}}{\Omega^{*}}+\frac{i\hat F_{ba}}{\Omega},
\end{aligned}
\end{equation}
Replacing the $\hat\sigma_{bc}$ in \eqref{eq:sigmabaeliminated} by the above formula, and keep the leading terms, we obtain Eq.(4-5) in the main text. However, after the optical field thoroughly passed the atom ensemble, the noise fields which broaden the atom state should be added to the $\hat\epsilon$ for compensating the Bosonic commutation relation. The optical field operator can be written, after passing the atom ensemble (after time $T$):
\begin{equation}\label{eq:opticalevolution}
\begin{aligned}
\hat{\varepsilon}(k,T)&=\hat{\varepsilon}(k,0){\rm exp}\left\{i\int_{0}^{T}\mathrm{d}t[kv-k^3c^3D(1-\eta)^3]\right\}\\&{\rm exp}\left\{\int_{0}^{T}\mathrm{d}t\left[A-{k^2c^2(1-\eta)^2}C\right]\right\}+\text{noise terms},
\end{aligned}
\end{equation}
where the first exponential is merely a phase shift (defined as $\Phi_T(k)$) and it is the second term which is an attentuation factor (defined as $\chi(k)$). Following Caves's approach, the above formula can be phenomologically written as:
\be\label{eq:Caves}
\hat{\epsilon}(k,T)=\chi(k) e^{i\Phi_T(k)}\hat{\epsilon}(k,0)+\sqrt{1-\chi(k)^2}\hat n_k,
\ee
where $\hat n$ describe the noise corresponds to all possible losses.

For a single photon state, the output optical quantum state now is ($\hat U$ is the evolution operator):
\be
\begin{split}
&|\Psi\rangle_{\rm out}=\hat U\int dx f_{\rm in}(x)\hat \epsilon_{x}^\dag|{\rm vac}\rangle=\hat U\int dk f_{\rm in}(k)\hat \epsilon_{k}^\dag|{\rm vac}\rangle\\
&=\int dkf_{\rm in}(k,T) [\chi^*(k) e^{-i\Phi_T(k)}\hat{\epsilon}_k^\dag(0)+\sqrt{1-|\chi(k)|^2}\hat n^\dag_k]|\rm vac\rangle.
\end{split}
\ee
where we have used the fact that $\hat U|\rm vac\rangle=|\rm vac\rangle$ and the $|{\rm vac}\rangle$ should be the joint Hilbert space of optical qubit and the noise field. In a more phenomenological way, it can be written as (for $R/L-$ polarisation respectively):
\be
\begin{split}
|\Psi\rangle^L_{\rm out}=C^{oL}_{1}|1_{\rm out L}\rangle_q\otimes|0\rangle^L_n+C^{oL}_{2}|0\rangle\otimes|1\rangle^L_n,\\
|\Psi\rangle^R_{\rm out}=C^{oR}_{1}|1_{\rm out R}\rangle_q\otimes|0\rangle^R_n+C^{oR}_{2}|0\rangle\otimes|1\rangle^R_n,
\end{split}
\ee 
with
\be
C^{oL/R}_1=\sqrt{\int dk |f_{\rm in}(k,T)|^2|\chi^{L/R}(k)|^2},
\ee
and $|C^{oL}_1|^2+|C^{oR}_2|^2=1$, which is the normalisation condition.

Since the optical qubit here is a single-photon state, we can use post-selection detections scheme for many applications. In the post-selection scheme, for the superposition input state $|\Psi\rangle=a|\Psi\rangle^L_{\rm in}+b|\Psi\rangle^R_{\rm in}$, we have post-selected output qubit state as:
\be
|\Psi\rangle_{\rm out}=aC^{oL}_{1}|1_{\rm out L}\rangle_q+bC^{oR}_{1}|1_{\rm out R}\rangle_q,
\ee
If we have balanced EIT storage process for the optical qubits with left and right polarisations, that is, $C^{oL}_1=C^{oR}_1$, we then have:
\be
|\Psi\rangle_{\rm out}=C^{oL}_{1}(a|1_{\rm out L}\rangle_q+b|1_{\rm out R}\rangle_q)
\ee
which perfectly matches the ideal qubit state. Therefore, practically, we can increase the loss of certain channels to balance these two EIT storage process, at the price of sacrifying the post selection probability.

\section{Cavity quantum electrodynamics:}
Microwave (MW) qubit is processed by a cavity QED system during the optical storage. In the above EIT process, we use single photon as optical qubits, therefore at most one atom populates on the Rydberg state $\left\vert{r_1}\right\rangle$. In addition, the wavelength of MW qubit is much larger than the size of Rydberg atom ensemble. 

In this case, a MW qubit enters the cavity and reflect from the cavity. In the rotating frame, the Hamiltonian is (the microwave cavity is on resonance with the transition $\left\vert{r_1}\right\rangle\rightarrow\left\vert{r_2}\right\rangle$):
\be
\begin{aligned}
H=\hbar g_m\hat{a}_c\hat{\sigma}_{r_2 r_1}+\hbar\sqrt{\kappa}\hat{a}_c\hat{b}^{\dagger}(0)+\hbar\sqrt{\kappa}_s\hat a_c\hat n^\dag+{\rm h.c.}
\end{aligned}
\ee
where $g_m$ is the coupling rate of the single atom to cavity field, $\kappa$ represents the cavity-qubit coupling strength, while $\kappa_s$ represent MW cavity loss rate.

The Heisenberg equations of motion for the cavity QED system reads:
\be
\begin{aligned}
&\dot{\hat{\sigma}}_{r_1 r_2}=-\gamma_{s}\hat{\sigma}_{r_1 r_2}+2ig_m\hat{\sigma}_{z}\hat{a}_{c}+\sqrt{2\gamma_s}\hat n_{\rm en},\\
&\dot{\hat{a}}_{c}=-ig_m\hat{\sigma}_{r_1 r_2}-\frac{\kappa+\kappa_s}{2}\hat{a}_c-\sqrt{\kappa}\hat{a}_{in}-\sqrt{\kappa}_s\hat n,\\
&\hat b_{\rm out}=-\hat b_{\rm in}+\sqrt{\kappa}\hat a_c,
\end{aligned}
\ee
where $\hat n_{\rm en}$ and $\hat n$ are noise induced by the decay of Rydberg state and cavity loss, respectively. These equations can be approximately treated in the strong coupling region and small cat-state amplitude (see Section D). In the Fourier domain, it has the following form:
\be
\begin{split}
&(-i\omega+\gamma_s)\hat{\sigma}_{r_1 r_2}(\omega)=-ig_m\hat{a}_{c}(\omega)+\sqrt{2\gamma_s}\hat n_{\rm en}(\omega),\\
&\left(-i\omega+\frac{\kappa+\kappa_s}{2}\right)\hat{a}_{c}(\omega)=-ig_m\hat{\sigma}_{r_1 r_2}(\omega)-\sqrt{\kappa}\hat{a}_{\rm in}(\omega)\\
&\qquad\qquad\qquad\qquad\qquad\quad-\sqrt{\kappa_s}\hat n(\omega).
\end{split}
\ee

Solving these equations in the Fourier domain, we have:
\be
\begin{split}
&\hat a_{c}(\omega)=\\
&\frac{\sqrt{\kappa}\hat b_{\rm in}(\omega)-\sqrt{\kappa_s}\hat n(\omega)+ig_m\sqrt{2\gamma_s}\hat n_{\rm en}(\omega)/(-i\omega+\gamma_s)}{i\omega+g_m^2/(i\omega-\gamma_s)
-(\kappa+\kappa_s)/2},
\end{split}
\ee
and the input-output relation is:
\be
\begin{split}
\hat b_{\rm out}(\omega)&=\frac{i\omega+g_m^2/(i\omega-\gamma_s)-(\kappa_s-\kappa)/2}
{i\omega+g_m^2/(i\omega-\gamma_s)-(\kappa_s+\kappa)/2}\hat b_{\rm in}(\omega)\\
&-\frac{\sqrt{\kappa\kappa_s}}{i\omega+g_m^2/(i\omega-\gamma_s)-(\kappa_s+\kappa)/2}\hat n(\omega)\\
&-\frac{ig_m\sqrt{2\kappa\gamma_s}/(i\omega-\gamma_s)}{i\omega+g_m^2/(i\omega-\gamma_s)-(\kappa_s+\kappa)/2}\hat n_{\rm en}(\omega).
\end{split}
\ee
The second and third term corresponds to the noise injection from the cavity loss and the Rydberg state decay. Using Cave's approach\,\cite{Caves1982}, we can phenomenologically write a noise term $\propto \hat n_{\rm in}$ which represents both these two contributions as:
\be\label{eq:binout}
\hat b_{\rm out}(\omega)=C_1(\omega)\hat b_{\rm in}(\Omega)+C_2(\omega)\hat n_{\rm in}(\omega),
\ee
where $\hat n_{\rm in}$ represents all possible noise contributions and 
\be\label{eq:MWCcoefficients}
\begin{split}
&C_1(\omega)=\frac{i\omega+g_m^2/(i\omega-\gamma_s)-(\kappa_s-\kappa)/2}
{i\omega+g_m^2/(i\omega-\gamma_s)-(\kappa_s+\kappa)/2},\\
&C_2(\omega)=\frac{i\omega-\gamma_{\rm eff}}{i\omega-\gamma_s}\frac{\sqrt{\kappa\kappa_s}}{i\omega+g_m^2/(i\omega-\gamma_s)-(\kappa_s+\kappa)/2},\\
&|C_1(\omega)|^2+|C_2(\omega)|^2=1,
\end{split}
\ee
with $\gamma_{\rm eff}=\sqrt{\gamma_s^2+2g_m^2\gamma_s/\kappa_s}$.

Let us consider that the input state is a Schroedinger cat state:
\be
\begin{split}
&|\Psi\rangle^{\rm MW}_{\rm in}=\\
&e^{-\alpha^2/2}\left[{\rm exp}\left(\alpha\int dx f_{\rm in}(x,t)\hat b^\dag_{\rm in}(x)\right)+(\alpha\rightarrow-\alpha)\right]|0\rangle_{qn},
\end{split}
\ee
where $|0\rangle_{qn}=|0\rangle_q\otimes|0\rangle_n$ is the joint vacuum state of MW qubit and the noise field.
The MW qubit-noise joint state evolution can be written as (we ignore the normalisation factor is ignored for brievty):
\be
\begin{split}
&|\Psi\rangle_{\rm out}=\hat U |\Psi\rangle^{\rm MW}_{\rm in}\\
&=\left[{\rm exp}\left(\alpha\int dx f_{\rm in}(x,t)\hat b^\dag_{\rm out}(x)\right)+(\alpha\rightarrow-\alpha)\right]|0\rangle_{qn}\\
&=|\Psi_\alpha\rangle_{\rm out}^q\otimes|E_\alpha\rangle+|\Psi_{-\alpha}\rangle_{\rm out}^q\otimes|E_{-\alpha}\rangle,
\end{split}
\ee 
where
\be
\begin{split}
&|\Psi_{\pm\alpha}\rangle_{\rm out}^q={\rm exp}\left[\pm\alpha\int d\omega f_{\rm in}(\omega,t)[C_1(\omega)\hat b^\dag_{\rm in}(\omega)\right]|0\rangle_{q}\\
&|E_{\pm\alpha}\rangle={\rm exp}\left[\pm\alpha\int d\omega f_{\rm in}(\omega,t)C_2(\omega)n^\dag_{\rm in}(\omega)\right]|0\rangle_n.
\end{split}
\ee


\section{Linearisation of cQED dynamics}
In the above derviation, we approximate the cQED dynamics by linearisation. Having a linearised dynamics of cQED system is also essential for a high fidelity processing of the MW Schroedinger cat qubit, otherwise, the nonlinear dynamics of cQED system will distort the MW Schroedinger cat qubit. In this section, two important points of this linear approximation are discussed.

Firstly, we discuss the validity conditions of this linearised dynamics: (1) the coherent amplitude $\alpha$ in the cat state can not be too large (but also can not be too small for application on quantum information science, usually $\alpha^2\geq 2$); (2) strong coupling region: $g_m^2/\kappa\gamma_s \gg 1$. The errors of the fidelity calculated by linearised cQED dynamics and the simulation result without linearisation have been shown in Fig.\,\ref{fig:linearisation}. It is clear that the validity of linear approximation $\langle \hat \sigma_z\rangle\approx 1/2$ will be degraded by decreasing the coupling strength $g_m$ and increasing the cavity bandwidth $\kappa$ and the coherent amplitude $\alpha$. In the following, we give have a detailed analyis of the validity of this linear approximation.

Our approach to examine the validity of linear approximation is to check the consistency of this approximation by substituting the result calculated under $\langle\hat\sigma_z\rangle\approx 1/2$ to the $\hat \sigma_z$-equation of motion:
\be
\dot{\hat {\sigma}}_z(t)=ig_m\hat \sigma_{r_1r_2}(t)\hat{a}_c^{\dagger}(t)+h.c.
\ee
Here, by Fourier-transforming the result obtained in the Appendix C, we have:
\be
\begin{split}
\hat{\sigma}_{r_1r_2}(t)=\int\frac{ig_m\sqrt{\kappa}e^{-i\omega (t-t')}}{-\omega^2-i(\kappa/2+\gamma_s)\omega+\kappa \gamma_s/2+g_m^2} \hat{a}_{\rm in}(t')d\omega dt'.
\end{split}
\ee
In our system, we can neglect the $\omega^2$ term as an approximation since we have long pulse as initial state:
\be
\hat{\sigma}_{r_1r_2}(t)=\frac{ig_m\sqrt{\kappa}}{(\kappa/2+\gamma_s)}\int\frac{e^{-i\omega (t-t')}}{-i\omega+\Gamma} \hat{a}_{\rm in}(t')d\omega dt',
\ee
with
\be
\Gamma=\frac{\kappa \gamma_s/2+g_m^2}{(\kappa/2+\gamma_s)}.
\ee
Solving the integral leads to (where $\hat \sigma_{r_1r_2}(0)=0$ is assumed):
\be
\hat{\sigma}_{r_1r_2}(t)=\frac{ig_m\sqrt{\kappa}}{\gamma_s+\kappa/2}\int_{-\infty} ^{t}e^{-\Gamma(t-t')}\hat{a}_{\rm in}(t')dt'.
\ee
The intracavity MW field $\hat a(t)$ is given by:
\be
\begin{split}
&\hat{a}_c(t)=\int\frac{\sqrt{\kappa}(i\omega-\gamma_s)e^{-i\omega(t-t')}}{-\omega^2-i(\gamma_s+\kappa/2)\omega+g_m^2+\kappa \gamma_s/2}\hat{a}_{\rm in}(t')d\omega dt'\\
&=\sqrt{\kappa}\int^t_{-\infty} e^{-(\gamma_s/2+\kappa/4)(t-t')}[\mathcal{\eta}\sin{\chi(t-t')}\\
&\qquad\qquad\qquad\qquad\qquad\quad-\cos{\chi(t-t')}]\hat{a}_{\rm in}(t')dt',
\end{split}
\ee
with
\be
\begin{split}
&\mathcal{\chi}=\sqrt{-(\gamma_s+\kappa/2)^2+4(g_m^2+\kappa\gamma_s/2)}/2,\\
&\mathcal{\eta}=(\kappa/2-\gamma_s)/\sqrt{-(\gamma_s+\kappa/2)^2+4(g_m^2+\kappa\gamma_s/2)}.
\end{split}
\ee
For an orders of magnitude estimation, let us assume that the input field is a steady coherent field $|\alpha\rangle$, then the above two integrals can be explicitly calculated and we have:
\be
\begin{split}
\langle\hat{\sigma}_{r_1r_2}\rangle=\frac{ig_m\sqrt{\kappa}\alpha}{(\gamma_s+\kappa/2)\Gamma},\quad
\langle\hat{a}_c\rangle=\frac{-2\sqrt{\kappa}\alpha\gamma_s}{2g_m^2+\kappa\gamma_s}.
\end{split}
\ee
It is clear that cavity field is very small for typical parameters, this reflects the fact that when the Rydberg state is occupied, the MW fields feel a reflection almost without phase delay. Then, to the leading order, the correction to $\langle \hat \sigma_z\rangle\approx 1/2$ is given by:
\be
\langle\delta\dot{\hat {\sigma}}_z\rangle=\frac{2\kappa g_m^2\gamma_s}{(g_m^2+\kappa\gamma_s/2)^2}\alpha^2_{\rm in}\approx \frac{2\kappa\gamma_s}{g_m^2}\alpha^2_{\rm in}.
\ee
Clearly, when we have large enough ``cooperativity'' $g_m^2/\kappa\gamma_s$ and small enough $\alpha$, the error to the linear approximation can be tiny.  This result has a transparent physical explanation, since larger cooperativitiy means stronger Purcell effect, which tends to de-excite the atom state $|r_2\rangle$. 

\begin{figure*}[h]
\centering
\includegraphics[width=.9\textwidth]{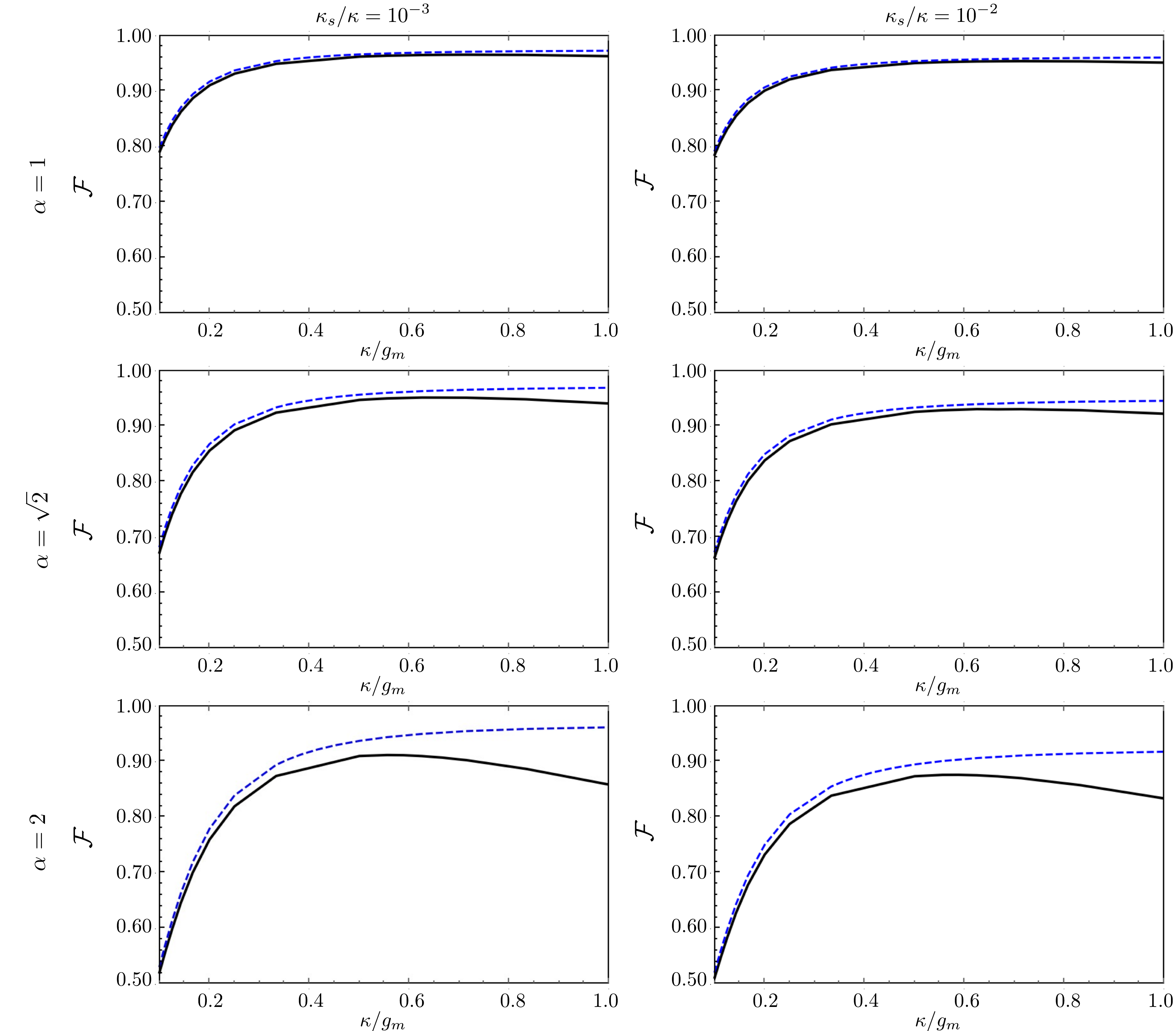}
\caption{The errors of fidelity calculation using linear approximation $\langle \hat \sigma_z\rangle\approx -1/2$. The dashed and solid curve are the results under linear approximation and the results obtained by full simulations, respectively. The left/right panel corresponds to $\kappa_s=0.001\kappa$ and $\kappa_s=0.01\kappa$, respectively. The three rows corresponds to $\alpha_0=1,\sqrt{2}$ and $2$, respectively. It is clear that the error increases with the increasing of $\kappa/g_m$ and $\alpha$, which is well consistent with our theretical analysis.}\label{fig:linearisation}
\end{figure*}

Secondly, we discuss the physical scenarios of this linearised dynamics. For example, when the input optical qubit is $(|1_L\rangle+|1_R\rangle)/\sqrt{2}$, the spin wave can be recorded in the superposition of $|r_1\rangle-|g\rangle$ and $|g'\rangle-|g\rangle$. This means that the Rydberg state is either being occupied at $|r_1\rangle$ or simply empty. Therefore, as long as the small $\alpha$ and large cooperativity conditions are satisfied, the linear approximation $\langle\hat\sigma_z\rangle=1/2$ can be applied to both of these two situations. As an example, suppose we have joint input state $|\psi\rangle=(|1_L\rangle+|1_R\rangle)/\sqrt{2}\otimes|\alpha\rangle$, the evolution can be written (in the lossless case) as:
\be
\begin{split}
\hat U|\psi\rangle=&\frac{1}{\sqrt{2}}|1_L\rangle\otimes \hat U{\rm exp}\left[\alpha\int dx f_{\rm in}(x)\hat b^\dag_{\rm in}(x)\right]|0\rangle\\
&+\frac{1}{\sqrt{2}}|1_R\rangle\otimes \hat U{\rm exp}\left[\alpha\int dx f_{\rm in}(x)\hat b^\dag_{\rm in}(x)\right]|0\rangle\\
=&\frac{1}{\sqrt{2}}|1_L\rangle\otimes {\rm exp}\left[\alpha\int dx f_{\rm in}(x)\hat b^{\dag g_m\neq0}_{\rm out}(x)\right]|0\rangle\\
&+\frac{1}{\sqrt{2}}|1_R\rangle\otimes {\rm exp}\left[\alpha\int dx f_{\rm in}(x)\hat b^{\dag g_m=0}_{\rm out}\right]|0\rangle,\\
\end{split}
\ee
where $\hat b_{\rm out}^{g_m=/\neq 0}$ is given in Eq.\,\eqref{eq:binout} with $g_m=0$ and $g_m\neq0$ corresponds to empty or occupied cavity, respectively. This method is used in computing the fidelity of our system.

\section{Fidelity}
For those transformations in the truth table of which no entanglement was generated, the gate fidelity can be simply calculated as $\mathcal{F}=\mathcal{F}_{\rm MW}\times\mathcal{F}_{\rm opt}$ and the result has been shown in the truth table in the main text.

However, a truth table can be simulated by the classical computer. For a quantum gate, there exists transformations of the CZ gate which generate optical-MW qubit entanglement. For example: $(|1\rangle_L+|1\rangle_R)\otimes(|\rm even\rangle+|\rm odd\rangle)\rightarrow |1\rangle_L\otimes(|\rm even\rangle+|\rm odd\rangle)+|1\rangle_R\otimes(|\rm even\rangle-|\rm odd\rangle)$. It is of crucial importance to consider the fidelity of these entanglement generation processes. We plot the resultant fidelity in Fig.\,\ref{fig:Fidelity_S} for two different MW cavity losses $\kappa_s/\kappa=0.01,0.001$ and different coherent amplitude of the Schroedinger cat state $\alpha=(1,\sqrt{2},2)$, where we use the rotation operators on the Bloch sphere (see the main text) to represent different optical and MW input states (see Fig.\,\ref{fig:Fidelity_S}). The operation of the quantum CZ gate on the separable input states can produce entangled states. In the following we will briefly discuss the way we compute the fidelity.
\\

The separable input qubit state:
\be
|\psi\rangle^{\rm in}_{q}=(c_1|1\rangle_L+c_2|1\rangle_R)\otimes (c_3|{\rm even}\rangle+c_4|\rm odd\rangle).
\ee
where 
\be
\begin{split}
&|\rm even\rangle=\frac{1}{\sqrt{2+e^{-2\alpha^2}}}(|\alpha\rangle+|-\alpha\rangle),\\
&|\rm odd\rangle=\frac{1}{\sqrt{2-e^{-2\alpha^2}}}(|\alpha\rangle-|-\alpha\rangle).
\end{split}
\ee
Then the output state \emph{after the post selection} is:
\be
\begin{split}
|\Psi\rangle^{\rm out}&=A_+C_1^{oL}|1_{\rm out L}\rangle_1|\alpha_{\rm out L}\rangle\otimes |g\rangle|E_{\alpha \rm L}\rangle\\
&+A_-C_1^{oL}|1_{\rm out L}\rangle_1|-\alpha_{\rm out L}\rangle\otimes |g\rangle|E_{-\alpha \rm L}\rangle\\
&+B_+C_1^{oR}|1_{\rm out R}\rangle_1|\alpha_{\rm out R}\rangle\otimes |g\rangle|E_{\alpha \rm R}
\rangle\\
&+B_-C_1^{oR}|1_{\rm out R}\rangle_1|-\alpha_{\rm out R}\rangle\otimes |g\rangle|E_{-\alpha \rm R}\rangle,
\end{split}
\ee
where we have omit those terms where the optical qubit is lost to the environment and 
\be
\begin{split}
&A_\pm=\frac{c_1c_3}{\sqrt{2+e^{-2\alpha^2}}}\pm\frac{c_1c_4}{\sqrt{2-e^{-2\alpha^2}}},\\
&B_\pm=\frac{c_2c_3}{\sqrt{2+e^{-2\alpha^2}}}\pm\frac{c_2c_4}{\sqrt{2-e^{-2\alpha^2}}}.
\end{split}
\ee
The ideal output field is given by:
\be
|\psi\rangle^{\rm ideal}_q= \sum_\pm A_\pm|1\rangle_L\otimes|\pm\alpha\rangle+B_\pm|1\rangle_R\otimes|\pm\alpha\rangle.
\ee

The output qubit density matrix after post-selection is: 
\be
\rho^{\rm out}_q={\rm Tr}_{g,E}[|\Psi\rangle^{\rm out}\langle\Psi|^{\rm out}],
\ee
where we trace out the atom and environmental degrees of freedom. Thus the fidelity is given by:
\be
\mathcal{F}={\rm Tr}[\hat \rho^{\rm out}_q\hat \rho_{\rm ideal}].
\ee

The result is quite cumbersome, and can be expressed as:
\be
\begin{split}
\mathcal{F}=&\sum_{ij,mn}^{ab,cd}f_{ij}f_{mn}f_{ab}f_{cd}\langle E_{mn}|E_{ij}\rangle\langle a_{\rm out}|1_{{\rm out} i}\rangle\langle 1_{{\rm out} m}|c_{\rm out}\rangle\\
&\langle (-1)^a b_{\rm out}|(-1)^i 1_{{\rm out} j}\rangle\langle (-1)^m 1_{{\rm out} n}|(-1)^c d_{\rm out}\rangle,
\end{split}
\ee
where the indices $(i,m,a,c)$ represent the polarisations $(L, R)$ and $(j,n,b,d)$ represent $(\alpha,-\alpha)$, and we define $(-1)^{R/L}=(-/+)1$. The $f-$coefficient is the matrix element of $\mathbb{F}$:
\be
\mathbb{F}=
\begin{pmatrix}
A_+\\
A_-\\
B_+\\
B_-\\
\end{pmatrix}\otimes
\begin{pmatrix}
A_+^*&A_-^*&B_+^*&B_-^*
\end{pmatrix},
\ee
where the row/column order is $(L,R,\alpha,-\alpha)$.

In computing the above fidelity, we need to use the following inner product of the MW environment states:
\be
\langle E_{\alpha i}|E_{-\alpha j}\rangle={\rm exp}\left[-\alpha^2\int d\omega|f_{\rm in}(\omega)|^2C^*_{2j}(\omega)C_{2i}(\omega)\right].
\ee
where $C_{2i}(\omega)$ is given by Eq.\,\eqref{eq:MWCcoefficients} with $g_m=/\neq 0$ when $i=R/L$, respectively.

As a special case, the fidelity for the truth table can be easily calculated as follows. The optical part $\mathcal{F}_{\rm opt}$, with the ideal output field:
\be
|\Psi\rangle^{\rm ideal}_{\rm out}=\int dk f_{\rm ideal}(k)\hat{\epsilon}_k^\dag(0)|{\rm vac}\rangle,
\ee
can be written as:
\be
\begin{split}
&\mathcal{F}_{\rm opt}=|\langle \Psi^{\rm ideal}_{\rm out}|\Psi_{\rm out}\rangle|^2\\
&\qquad=\int dk f_{\rm ideal}^*(k')f_{\rm in}(k,T) \chi^*(k) e^{-i\Phi_T(k)}.\\
\end{split}
\ee

The MW part $\mathcal{F}_{\rm MW}$, is given by:
\be
\begin{split}
\mathcal{F}&=\langle \Psi|_{\rm in}^{\rm MW}\hat \rho^q_{\rm out}|\Psi\rangle_{\rm in}^{\rm MW}\\
&=4\mathcal{N}(1+\xi)(\cos{2\alpha^2\Lambda_i}+\cosh{2\alpha^2\Lambda_r}),
\end{split}
\ee
where we have:
\be
\begin{split}
&\Lambda=\int d\omega |f_{\rm in}(\omega)|^2C^*_{1}(\omega)=\Lambda_r+i\Lambda_i,\\
&\xi=\langle E_\alpha|E_{-\alpha}\rangle={\rm exp}\left[-\alpha^2\int d\omega|f_{\rm in}(\omega)C_2(\omega)|^2\right].
\end{split}
\ee
and the normalisation factor:
\be
\mathcal{N}={\rm exp}[-2\alpha_0^2]/(2\pm {\rm exp}[-2\alpha_0^2])^2.
\ee
The final fidelity for the truth table is simply $\mathcal{F}=\mathcal{F}_{\rm MW}\times\mathcal{F}_{\rm opt}$.

\begin{figure*}[h]
\centering
\includegraphics[width=0.8\textwidth]{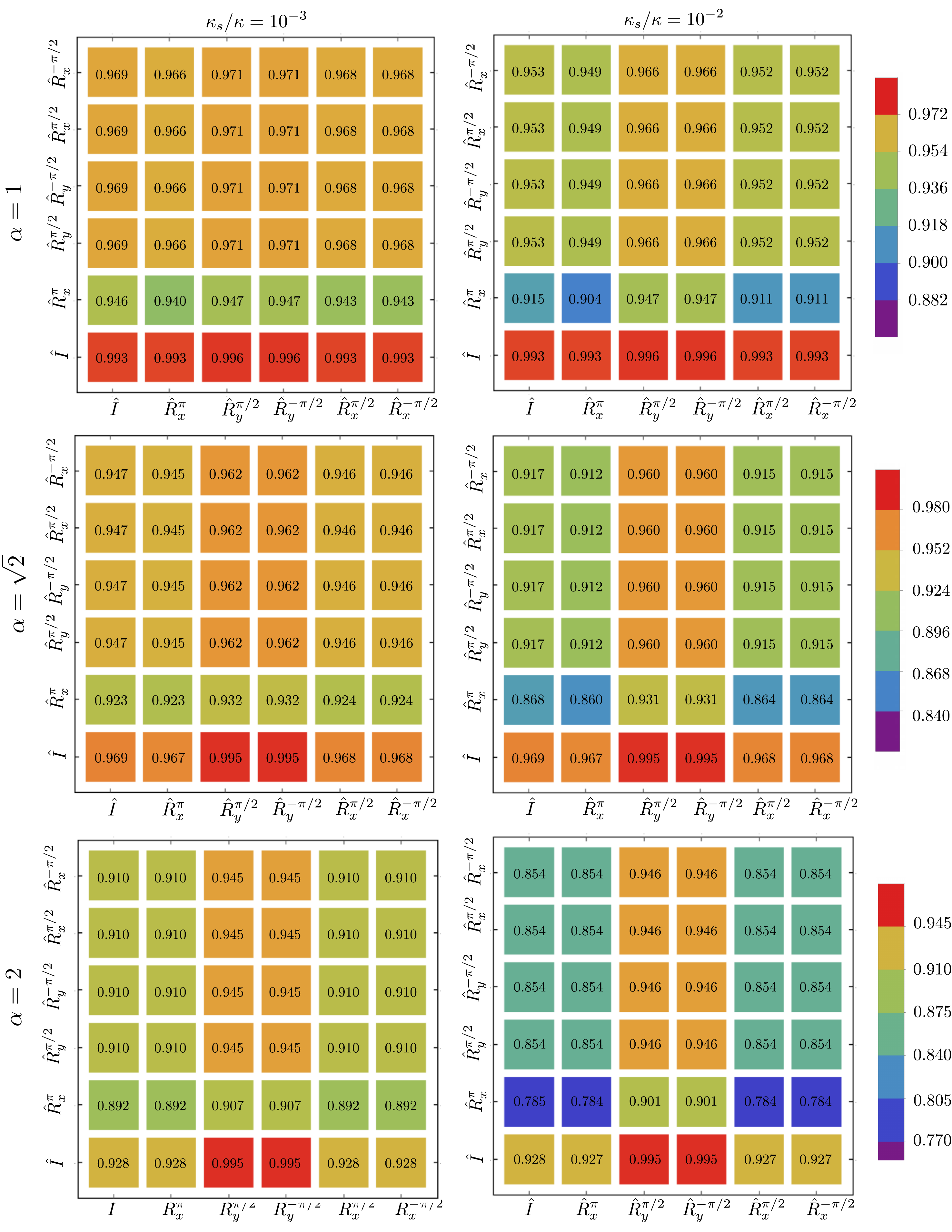}
\caption{The fidelity of the hybrid quantum gate for CZ gate with separable input states $|\psi\rangle^o_{q}\otimes|\psi\rangle^{\rm MW}_q$ when $\kappa_s=10^{-3}\kappa$ and $\kappa_s=10^{-2}\kappa$. The vertical and horizontal axis represent optical qubit states $|\psi\rangle^o_{q}$ and the MW qubit states $|\psi\rangle^{\rm MW}_q$, respectively. The color represent the fidelity values. For illustrative purpose, we choose a fiducial state as $|\psi_{\rm FD}\rangle=|L\rangle$ or $|\rm even\rangle$, and other states are represented by operations on the Bloch sphere (see the main text) $\hat R_j^\theta$ where $j$ is the axis index and $\theta$ is the rotation angle. The concrete fidelity value of each input state is also marked.}\label{fig:Fidelity_S}
\end{figure*}

\section{parameters}
In choosing parameters, the strong interaction strength of cavity QED system leads to some constraints to other system parameters. The Rabi frequency of cQED system driven by a single photon in continuous mode is given by:
\be
g_m=\sqrt{\frac{\hbar\omega_c}{2\epsilon_0 V_c}}\frac{p_{r_1r_2}}{\hbar}\frac{u(r)}{\kappa},
\ee
where $p_{r_1r_2}$ is the Rydberg dipole transition amplitude, $u(r)$ is the cross-sectional field distribution and $\kappa$ is the cavity bandwidth. To achieve a strong $g_m$, the $\kappa$ can not be too large, which means the cavity damping time can not be too short. At the same time, for a high fidelity gate, the life time of atom state $|r_1\rangle$ can not be shorter than the cavity damping time, that is, $\gamma_{r_1g}<\kappa$ is required.

\widetext{
\begin{table}[h]
        \centering
       \begin{tabular}{|c|c|c|}
         \hline
         Rydberg state 1&{$\vert {r_1}\rangle$} & {$\vert {69S_{1/2},F=m_F=2}\rangle$} \\
         \hline
         Rydberg state 2&{$\vert {r_2}\rangle$} & {$\vert {69P_{3/2},F=m_F=3}\rangle$} \\
         \hline
         cQED coupling strength &{$g_m$} & {$2\pi\times2.723\,\rm MHz$} \\
         \hline
         Decay rate of Rydberg state 1&$\gamma_{r_1g}$  &  $2\pi\times 3.50\,\rm kHz$  \\
         \hline
         Decay rate between Rydberg states&$\gamma_{r_2r_1}$  &  $2\pi \times4.78\,\rm kHz$  \\
         \hline
         MW cavity bandwidth& $\kappa$  &  $2\pi\times 2$\,MHz\\
         \hline
         Optical pulse duration time&$T_0$  &    $0.5\mu s$ \\
         \hline
         EIT storage time &$T_{\rm EIT}$  &     $16\mu s$ \\
         \hline
         Decay rate of excited state &$\gamma_{eg}$  &  $ 2\pi\times 3\,\rm MHz$ \\
         \hline
         Optical cross-sectional diameter &$w$  &  $8\,\mu m$ \\
         \hline
         Superconducting circuit electrodes distance&$w_c$   &   $30\,\mu m$  \\
         \hline
         Atom number density &$\varrho$    &  $8\times 10^{11}\,\rm cm^{-3}$ \\
         \hline
         Length of atom ensemble &$L_{a}$ &  $0.4\rm mm$ \\
         \hline
         Vacuum Rabi frequency for right-polarised optical field &$g_R$   &  $ 2\pi\times 0.012\,\rm MHz$  \\
         \hline
          Vacuum Rabi frequency for left-polarised optical field&$g_L$   &   $2\pi \times 0.029\times\,\rm MHz$ \\
         \hline
       \end{tabular}
       \caption{Sample Parameters for the hybrid quantum gate}
\end{table}}

Reducing the temperature of the atomic ensemble can lead to small $\gamma_{r_1g}$, since the thermal Doppler effect of atoms mainly contributes to the $\gamma_{r_1g}$ as:
$\gamma_{r_1g}\approx(k_{ge}-k_{er_1})k_BT/m$. Since the single photon optical qubit at most excits only one atom to the Rydberg state, therefore the collective effect discussed in\,\cite{Saijun2020} has no enhancement to the $\gamma_{r_1g}$. If we bring the temperature down to $0.2 \mu K$. we will get $\gamma_{r_1g}=22$\,kHz.  Previous work for achieving the strong electric dipole moment of Rydberg atom proposed to use MW resonator based on superconducting circuits\,\cite{pritchard2014hybrid,petrosyan2009reversible,morgan2020coupling}. This method could be also useful for our proposal. 
We assume that the grounded superconducting electrodes locates at distance $w_c=30\mu m$, which confines the cavity field within the effective volume $V_c=\int\mathrm{d}^3r|u(r)|^2\approx\frac{\pi}{2}\omega^2L$ and the cavity length is taken to be $L=1.38$\,cm. The relevant transition dipole matrix element from $\vert r_1\rangle$ to $\vert r_2\rangle$ is $2847a_0e$. We also set the distance between atom ensemble and the coplanar superconducting circuit surface is determined by the mode function equation $u(r)\approx e^{-1}$. Then we can obtain $g_m=2\pi\times2.723$\,MHz. In Fig.\,\ref{fig:Fidelity_S}, it is clear that the fidelity depends on the value of $\kappa_s$. For coplanar superconducting resonators, recent state-of-art can achieve the loss level to $\kappa_s/\kappa=0.001$ (i.e. the internal quality factor to be $10^6-10^7$) where the loss is dominated by the interfacial and surface two-level-states (the so-called TLS loss)\,\cite{Cleland2012}. 

\end{document}